\documentclass[12pt,fleqn]{article}

\usepackage{graphicx}
\usepackage{pdftricks}
\usepackage[utf8]{inputenc}
\usepackage{longtable}
\usepackage{amsmath}
\RequirePackage{lineno}
\usepackage{epstopdf}

\begin{document}

\title{Thermodynamic Black Holes}

\author{George Ruppeiner\footnote{Division of Natural Sciences, New College of Florida, 5800 Bay Shore Road, Sarasota, FL 34243, USA (ruppeiner@ncf.edu)}}

\maketitle

\begin{abstract}
Black holes pose great difficulties for theory since gravity and quantum theory must be combined in some as yet unknown way. An additional difficulty is that detailed black hole observational data to guide theorists is lacking. In this paper, I sidestep the difficulties of combining gravity and quantum theory by employing black hole thermodynamics augmented by ideas from the information geometry of thermodynamics. I propose a purely thermodynamic agenda for choosing correct candidate black hole thermodynamic scaled equations of state, parameterized by two exponents. These two adjustable exponents may be set to accommodate additional black hole information, either from astrophysical observations or from some microscopic theory, such as string theory. My approach assumes implicitly that the as yet unknown microscopic black hole constituents have strong effective interactions between them, of a type found in critical phenomena. In this picture, the details of the microscopic interaction forces are not important, and the essential macroscopic picture emerges from general assumptions about the number of independent thermodynamic variables, types of critical points, boundary conditions, and analyticity. I use the simple Kerr and Reissner-Nordstr$\ddot{\mbox{o}}$m black holes for guidance, and find candidate equations of state that embody a number of the features of these purely gravitational models. My approach may offer a productive new way to select black hole thermodynamic equations of state representing both gravitational and quantum properties.
\end{abstract}

\noindent Keywords: Black hole thermodynamics; information geometry of thermodynamics; thermodynamic curvature; Kerr black hole; Reissner-Nordstr$\ddot{\mbox{o}}$m black hole; critical phenomena\\

\section{Introduction}

Thermodynamics rests on general principles spanning a wide range of physical systems, from pure fluids to black holes. I focus here on the theory of black holes that has been brought into thermodynamics via black hole thermodynamics (BHT) \cite{Wald2001}. The idea contributed in this paper is that there is basic thermodynamic information to be found by considering the interplay between macroscopic and mesoscopic size scales in a system. This viewpoint is applied by linking the thermodynamic curvature $R$ with the free energy $\phi$ from the theory of critical phenomena. Applications in conventional thermodynamic have yielded strong results. We might reasonably expect this method to extend to the black hole scenario.

\par
In the conventional thermodynamic scenario, thermodynamic fluctuation theory allows us to probe mesoscopic size scales using macroscopic information \cite{Landau1980, Pathria2011}. Perhaps not as well known is that the theory of critical phenomena, via hyperscaling, allows us to reverse the tables, and use mesoscopic properties to illuminate the macroscopic thermodynamics \cite{Goodstein1975}. These complementary ideas are combined here to get a differential equation for the free energy by using the thermodynamic Ricci curvature scalar $R$ from the thermodynamic information geometry. Assuming that the free energy follows a scaled form allows us to readily solve for the thermodynamics.

\par
\begin{figure}
\centering
\includegraphics[width=9cm]{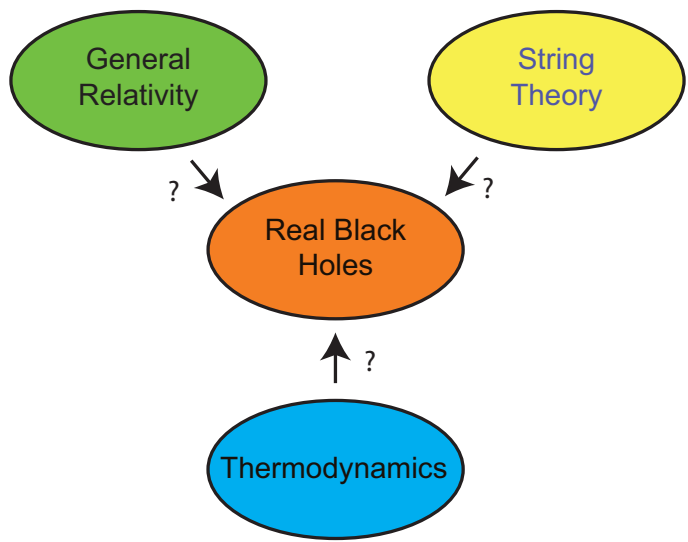}
\caption{Which theory correctly explains real black holes? General relativity lacks quantum mechanics, and string theory lacks consensus on which model correctly represents physical reality. I propose that a thermodynamic approach might help sort these issues out.}
\label{fig:Figure1}
\end{figure}

Much work has been done calculating the thermodynamic $R$ from known black hole models; see \cite{Sahay2017a} for a brief recent review. But this agenda is restricted in terms of its physical impact because the models in play typically originate either from general relativity or from string theory. There are good reasons for believing that relativity and quantum mechanics must be combined in some fundamental way in the black hole scenario, and how this is done is not clear. There is no consensus which of a number of string theory models give the correct physical results. These points are illustrated in Figure \ref{fig:Figure1}. The approach in this paper is more active. Instead of passively computing $R$ for models that may be inadequate for representing physical reality, I proceed with model independent thermodynamics.

\par
In describing the contribution of the thermodynamic information geometry to black hole thermodynamics, \r{A}man et al. \cite{Aman2015} expressed this opinion: ``There is a dream, however, which is that [thermodynamic information geometry] may have something to say about the obvious question: what properties must a function have if it is to serve as an entropy function for black holes?'' My hope here is that my purely thermodynamic approach offers the opportunity for writing down correct classes of thermodynamic black holes, whose comparison with what is known can offer guidance in the search for a correct theory.

\section{Thermodynamic information geometry}

In this section, I summarize the thermodynamic information geometry, starting with thermodynamic fluctuation theory.

\subsection{Thermodynamic fluctuation theory}

\par
The first element of thermodynamic fluctuation theory consists of the separation of the universe into two distinctive parts, a finite system and an environment tending to infinite size. System and environment interact via some boundary across which certain conserved parameters are allowed to fluctuate back and forth. This boundary serves only to separate the two parts, and has zero heat capacity. I consider only cases with two conserved fluctuating variables, but the formalism readily generalizes to more variables. A summary of how thermodynamic fluctuation theory is applied to BHT is given in \cite{Ruppeiner2007}.

\par
\begin{figure}
\begin{minipage}[b]{0.5\linewidth}
\hspace{0.8 in}
\includegraphics[width=1.5 in]{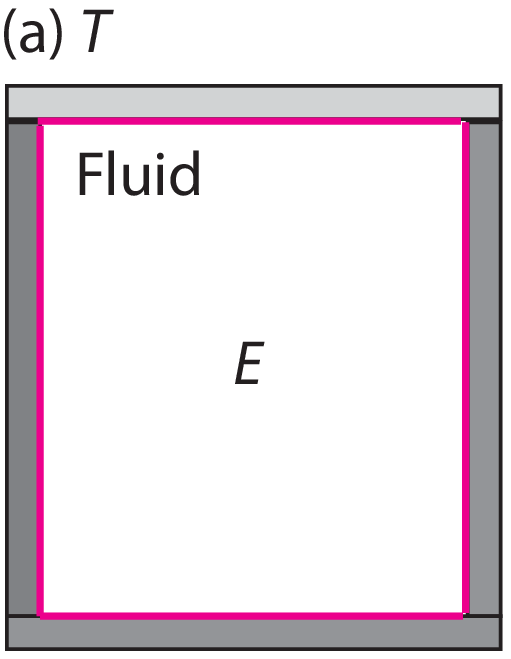}
\end{minipage}
\hspace{0.0 in}
\begin{minipage}[b]{0.5\linewidth}
\includegraphics[width=1.5 in]{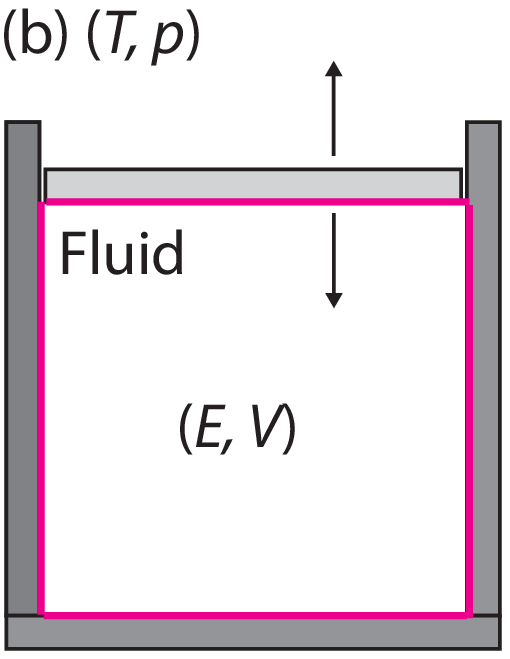}
\end{minipage}
\begin{minipage}[b]{0.5\linewidth}
\hspace{0.8 in}
\includegraphics[width=1.5 in]{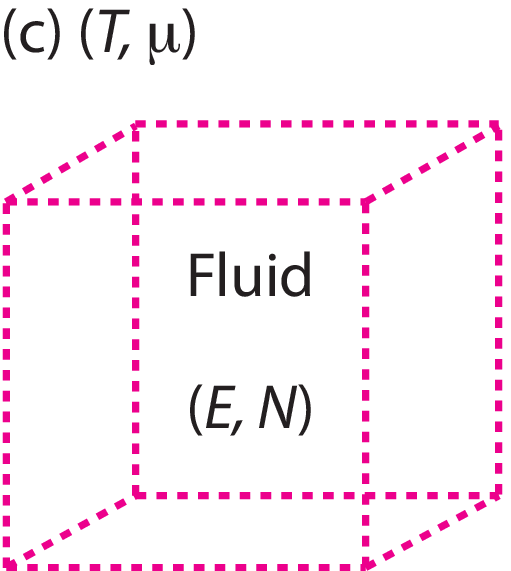}
\end{minipage}
\hspace{0.0 in}
\begin{minipage}[b]{0.5\linewidth}
\includegraphics[width=1.5 in]{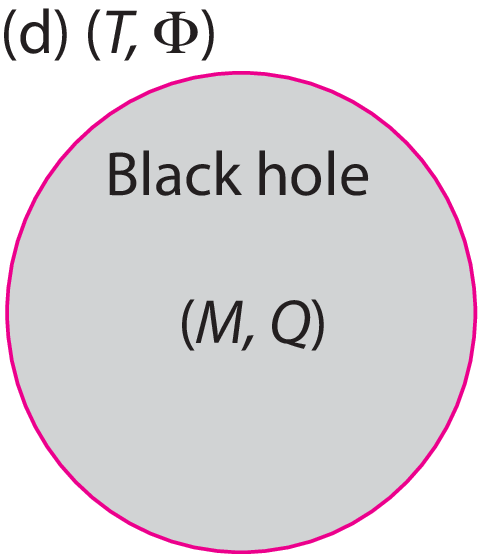}
\end{minipage}
\caption{Four thermodynamic systems, environments, and boundaries. Cases of conserved fluctuating variables are: $E$ or $M$ = energy, $V$ = volume, $N$ = number of particles, and $Q$ = charge. Also shown are conjugate nonfluctuating environment parameters: $T$ = temperature, $p$ = pressure, $\mu$ = chemical potential, and $\Phi$ = electric potential. We have: (a) a closed fluid system exchanging heat with its environment; (b) a closed fluid system exchanging heat and volume via a movable piston; (c) an open fluid system with fixed volume exchanging energy and particles; and (d) a Reissner-Nordstr$\ddot{\mbox{o}}$m black hole exchanging energy and charge via Hawking radiation at the event horizon boundary.}
\label{fig:Figure2}
\end{figure}

Let system and environment each have pairs of fluctuating parameters $(X^1, X^2)$ and $(X^1_e, X^2_e)$, respectively. They follow a conservation law

\begin{equation} X^{\alpha}+X^{\alpha}_{e} = X^{\alpha}_{tot}=\mbox{constant}, \label{10}\end{equation}

\noindent with the index $\alpha=1,2$. Figure \ref{fig:Figure2} shows examples of systems, environments, and boundaries in both the pure fluid and the black hole scenarios. I assume that the environments are always very large, and conventional in character. Namely, $(X^1_e, X^2_e)$ and the environmental entropy $S_e=S_e(X^1_e, X^2_e)$ all scale up in proportion to the environmental size.

\par
In this scenario, it turns out that the only properties of the environment relevant to the fluctuations of the system are the values of two intensive environment parameters

\begin{equation}F_{e\alpha}=\frac{\partial S_e}{\partial X_e^{\alpha}}\label{15}\end{equation}

\noindent around which the fluctuations take place, e. g., temperature, chemical potential $\ldots$.

\par
Let $S=S(X^1, X^2)$ denote the entropy of the system. I make no assumption that $S$ depends on its variables in the same way that $S_e$ depends on its. Namely, the system is not necessarily a scaled down version of the environment. The only case in Fig. \ref{fig:Figure2} where system and environment are necessarily of the same type is Fig. \ref{fig:Figure2}(c), where the boundary is imaginary.

\par
A frequent concern in the literature is the unconventional black hole scaling with system size. For example, combining two identical black holes does not result in a black hole with twice the volume. But system size scaling does not enter into the thermodynamic fluctuation theory described below. $S=S(X^1,X^2)$ is simply taken to be some function of its two variables.

\par
The total entropy of the universe is

\begin{equation}S_{tot}=S(X^1, X^2)+S_e(X^1_{e}, X^2_{e}).\label{20}\end{equation}

\noindent With fluctuations stable about some environment state $F_{e\alpha}$, the generalized second law of thermodynamics requires that the equilibrium state has a local maximum for $S_{tot}$. Let $\tilde{X}^{\alpha}$ be the values of $X^{\alpha}$ that maximize $S_{tot}$ [with the corresponding $\tilde{X}^{\alpha}_e$ given by Eq. (\ref{10})]. Expanding $S_{tot}$ to first order in 

\begin{equation}\Delta X^\alpha=X^{\alpha}-\tilde{X}^{\alpha}, \label{30}\end{equation}

\noindent and using the conservation laws Eq. (\ref{10}), yields

\begin{equation} \frac{\partial S}{\partial X^{\alpha}}=\frac{\partial S_e}{\partial X^{\alpha}_e}. \label{40}\end{equation}

\noindent This pair of algebraic equations equating conjugate variables may be solved for $\tilde{X}^{\alpha}$ if necessary.

\par
Expanding $S_{tot}$ in Eq. (\ref{20}) to second order about its local maximum, and using Eq. (\ref{10}), yields 

\begin{equation}\Delta S_{tot}=\frac{1}{2}\frac{\partial^2 S}{\partial X^{\mu}\partial X^{\nu}}\Delta X^{\mu} \Delta X^{\nu} + \frac{1}{2}\frac{\partial^2 S_e}{\partial X^{\mu}_e \partial X^{\nu}_e}\Delta X^{\mu} \Delta X^{\nu}.\label{50}\end{equation}

\noindent Since both $X^{\alpha}_e$ and $S_e$ scale up in proportion to the size of the environment (which is assumed to trend to infinity) the second term on the right-hand side of Eq. (\ref{50}) is much smaller than the first, and

\begin{equation}\Delta S_{tot}=\frac{1}{2}\frac{\partial^2 S}{\partial X^{\mu}\partial X^{\nu}}\Delta X^{\mu} \Delta X^{\nu}.\label{60}\end{equation}

\noindent Clearly, variations in $S_{tot}$ depend only on the detailed thermodynamic properties of the {\it system}, which are embodied in the Hessian elements of $S$. Effects of the environment appear just through the values of the derivative $F_{e\alpha}$ that set the state about which the system fluctuates, and where the Hessian elements in Eq. (\ref{60}) get evaluated.

\par
Thermodynamic fluctuation theory \cite{Landau1980, Pathria2011} has the probability of the system fluctuation as

\begin{equation}P dX^1 dX^2 \propto\exp\left(\frac{\Delta S_{tot}}{k_B}\right)dX^1 dX^2, \label{80}\end{equation}

\noindent where $k_B$ is Boltzmann's constant. Eq. (\ref{60}) now leads to 

\begin{equation}P\propto\exp\left(-\frac{1}{2}\,\Delta\ell^2 \right), \label{90}\end{equation}

\noindent where

\begin{equation}\Delta\ell^2=-\frac{1}{k_B}\frac{\partial^2 S}{\partial X^{\mu}\partial X^{\nu}}\Delta X^{\mu} \Delta X^{\nu}. \label{100}\end{equation}

\noindent $\Delta\ell^2$ is the line element in the thermodynamic information geometry described in the next subsection. Henceforth, I take $k_B=1$, which makes the entropy dimensionless.

\par
If $\Delta\ell^2$ is positive definite, we say that the fluctuations are stable. Necessary and sufficient conditions that $\Delta\ell^2$ be positive definite are that

\begin{equation}p_1=\frac{\partial^2 S}{\partial (X^{1})^2}\label{}\end{equation}

\noindent and

\begin{equation}p_2=\frac{\partial^2 S}{\partial (X^1)^2}\frac{\partial^2 S}{\partial (X^2)^2}-\left(\frac{\partial^2 S}{\partial X^1\partial X^2}\right)^2\label{}\end{equation}

\noindent both be positive.

\par
Thermodynamic stability is frequently lacking in self-gravitating scenarios, because of negative heat capacity. But, the widespread attitude in the literature has been not to worry excessively about this. Aman et al. \cite{Aman2003} put it like this: ``Nevertheless we believe that the Ruppeiner geometry of black holes is telling us something; our justification is mainly the {\it a posteriori} one that once it has been worked out for some examples we will find an interesting pattern.'' This is certainly my working philosophy in this paper.

\par
Frequent concern is expressed in the literature about the ``ensemble dependence" of BHT, but authors are not always clear about what they mean. The phrase ``ensemble dependence'' is used in ordinary statistical mechanics, where statistical ensembles over microstates are summed to get the thermodynamic properties. For finite systems these sums depend on the character of the boundary between the system and the environment. For example, Figs. \ref{fig:Figure2}(a), \ref{fig:Figure2}(b), and \ref{fig:Figure2}(c) yield the canonical, the Boguslavski, and the grand canonical ensembles, respectively. A central principle of statistical mechanics has the ensembles all yielding equivalent thermodynamic results for infinite systems (``the existence of the thermodynamic limit'').

\par
Since black holes are always of finite size, we might think that they must suffer from ``ensemble dependence,'' and that this points to some fundamental conflict with statistical mechanics. But there is no conflict. First, the general relativity black hole models have no underlying microscopic structures, so there are actually no microscopic statistical ensembles in play there. Second, and perhaps more to the point, in fluctuation theory each boundary between the system and the environment contributes its own character, with each leading to its own distinctive quadratic form in Eq. (\ref{60}). So, for example, the Reissner-Nordstr$\ddot{\mbox{o}}$m (RN) black hole in Fig. \ref{fig:Figure2}(d) has an event horizon admitting energy and charge, and the Kerr black hole substitutes angular momentum for charge. These are two different physical problems, and not merely some form of different ensemble representations of the same thing. These points are made carefully in \cite{Sahay2017a, Ruppeiner2007}.

\subsection{Thermodynamic information geometry}

\par
Motivated by Eq. (\ref{100}), we define the thermodynamic information metric elements as \cite{Ruppeiner1979, Ruppeiner1995, Diosi2000}

\begin{equation}g_{\alpha\beta}=-\frac{\partial^2 S}{\partial X^{\mu}\partial X^{\nu}}. \label{110}\end{equation}

\noindent This metric leads to the thermodynamic Ricci curvature scalar

\begin{equation} \begin{array}{lr} {\displaystyle R= -\frac{1}{\sqrt{g}} \left[ \frac{\partial}{\partial X^1}\left(\frac{g_{12}}{g_{11}\sqrt{g}}\frac{\partial g_{11}}{\partial X^2}-\frac{1}{\sqrt{g}}\frac{\partial g_{22}}{\partial X^1}\right) \right.} \\ \hspace{3.6cm} + {\displaystyle \left. \frac{\partial}{\partial X^2}\left(\frac{2}{\sqrt{g}} \frac{\partial g_{12}}{\partial X^1} -\frac{1}{\sqrt{g}}\frac{\partial g_{11}}{\partial X^2}-\frac{g_{12}}{g_{11}\sqrt{g}}\frac{\partial g_{11}}{\partial X^1}\right)\right],} \end{array} \label{120}\end{equation}

\noindent where $g=g_{11}g_{22}-g_{12}^2$. I use the sign convention of Weinberg \cite{Weinberg1972}, where $R$ for the two-sphere is negative.

\par
Numerous applications have demonstrated that $R$ gives information about the character of microscopic interaction forces; see \cite{Ruppeiner2010} for a brief recent review. Near the critical point $|R|$ is proportional to $\xi^d$:

\begin{equation} |R|\propto\xi^d,\label{123}\end{equation}

\noindent with $\xi$ the correlation length, and $d$ the spatial dimensionality. The sign of $R$ is positive/negative for microscopic interactions repulsive/attractive. The situation is particularly clear in the case of pure fluids \cite{Ruppeiner2012b, May2012, Ruppeiner2012a}.

\subsection{Hyperscaling and the geometric equation}

\par
A key ingredient from the theory of critical phenomena, one that completes the basic theory in this paper, is hyperscaling. Hyperscaling asserts that the singular part of the thermodynamic potential per volume $\phi$ is proportional to the inverse of the correlation volume \cite{Goodstein1975, Widom74}:

\begin{equation} \phi\sim\xi^{-d}.\label{125}\end{equation}

\noindent Combining this with Eq. (\ref{123}) leads to the geometric equation (GE):

\begin{equation}R=-\frac{\kappa}{\phi},\label{130}\end{equation}

\noindent where $\kappa$ is a dimensionless constant of order unity that the solution process always determines \cite{Ruppeiner1991}.

\par
This loose derivation is presented only to give the reader some idea of the source of the GE Eq. (\ref{130}). In practical applications, it is the GE that is important. Its precise expression is shaped more by mathematical consistency in the solution process that by any derivation. If the reader wants, he or she may regard the GE as a postulate spanning a range of systems, whose validity is tested by the quality of its results. The precise definition of the free energy $\phi$ is taken from applications in ordinary thermodynamics:

\begin{equation}\phi=S-F_1 X^1-F_2 X^2,\label{140}\end{equation}

\noindent where

\begin{equation}F_{\alpha}=\frac{\partial S}{\partial X^{\alpha}}.\label{145}\end{equation}

\noindent It has been argued that this GE extends in direct fashion to three or more independent variables \cite{Ruppeiner1998}. For a detailed recent discussion of the expression and the justification of the GE, see \cite{Ruppeiner2014, Ruppeiner2015}. These references also discuss the subtraction of a possible background term from $\phi$ (a subtraction not in play in this paper).

\par
In ordinary thermodynamics, the metric elements and $\phi$ were always divided by the fixed volume, to make them scale invariant. However, a per volume representation is necessary neither for the calculation nor for the physical interpretation of $R$. In BHT, there is generally no constant scaling factor to divide the metric elements by, so I simply omit such a division. Without it, $R$ is dimensionless, and its physical interpretation is that its magnitude is proportional to the number of correlated Planck area pixels on the event horizon \cite{Ruppeiner2008}.

\par
As is revealed by Eqs. (\ref{110}), (\ref{120}), (\ref{140}), and (\ref{145}), the GE may be written as a third-order partial differential equation (PDE) for $S$. [The fourth derivatives of $S$ subtract out in Eq. (\ref{120}).] Since both sides of Eq. (\ref {130}) are invariant scalar quantities, we could rewrite this equation in terms of other variables, yielding the same physical results. As I discuss below, I solve the GE only in cases where the entropy is assumed to be a generalized homogeneous function. This assumption simplifies the GE to a readily solvable third-order ordinary differential equation (ODE).

\par
The GE has been solved in several ordinary thermodynamic scenarios, with good results; see Table \ref{tab:table1}. Solutions were made in one or two analytic sections depending on whether or not there was a critical point at temperature other than zero. Expanding this list of solutions would further test the effectiveness of this method, and how reliable it might be in the black hole scenario.

\begin{table}[h!]
\caption{Tests of the GE in ordinary thermodynamics. $n$ denotes the number of independent thermodynamic variables, $d$ the spatial dimension, $a\#$ the number of analytic sections, and $f$ the number of fit parameters for the scaled equation.}
\label{tab:table1}
\centering
\begin{tabular}{lccccl}\\
\hline
\hline
System										& $n$	& $d$	& $a\#$	& $f$		& notes						\\
\hline
mean field theory \cite{Ruppeiner1991}				& $2$	& $-$	& $1$	& $0$	& exact			\\
critical point \cite{Ruppeiner1991}					& $2$	& $3$	& $2$	& $2$	& $\chi^2\sim 1$		\\
galaxy clustering \cite{Ruppeiner1996}				& $2$	& $3$	& $1$	& $0$	& qualitative		\\
corrections to scaling \cite{Ruppeiner1998}			& $3$	& $3$	& $2$	& $3$	& unclear			\\
ideal gas paramagnet \cite{Kaviani1999}				& $3$	& $3$	& $1$	& $0$	& exact		\\
power law interacting fluids \cite{Ruppeiner2005}		& $2$	& $3$	& $1$	& $1$	& qualitative		\\
unitary fermi fluid \cite{Ruppeiner2014, Ruppeiner2015}	& $2$	& $3$	& $2$	& $4$	& $\chi^2\sim 2$		\\
1D ferromagnetic Ising \cite{Ruppeiner2018}			& $2$	&$1$		& $1$	& $0$	& exact			\\
\hline
\hline
\end{tabular}
\end{table}

\subsection{Generalized homogeneous functions}

In this subsection, I discuss generalized homogeneous functions (GHF) of the form:

\begin{equation}\lambda\,S({X^1, X^2})=S(\lambda^{a_1} X^1,\lambda^{a_2} X^2),\label{150}\end{equation}

\noindent where $\lambda$ is a factor, and $a_1$ and $a_2$ are constant ``critical exponents.'' This is equivalent to writing \cite{Stanley1999}

\begin{equation}S({X^1, X^2}) = (X^1)^a\,Y\left[\frac{X^2}{(X^1)^{b}}\right],\label{160}\end{equation}

\noindent where $a=1/a_1$ and $b=a_2/a_1$. $Y(z)$ is a function of a single variable:

\begin{equation} z=\frac{X^2}{(X^1)^{b}}.\label{170}\end{equation}

\par It is easy to prove that if a function is a GHF, then both its derivatives are GHF's, and that multiplying, dividing, adding, or subtracting two GHF's results in a GHF. Hence, if $S({X^1, X^2})$ is a GHF, then the derived thermodynamic metric elements and the thermodynamic curvature are GHF's \cite{Ruppeiner2007}.

\par
GHF's are at the foundation of the modern theory of critical phenomena \cite{Stanley1999}. They were also central to solving the GE for the systems in Table \ref{tab:table1}. GHF's also appear in the Kerr and the RN black holes discussed in the next subsection. Although I focus in this paper on the extremal black hole limit, non-extremal black hole phase transitions occur as well. These BHT phase transitions usually have the van der Waals mean field theory exponents, with the RN-AdS model serving as the prototype \cite{Chamblin1999}. Potentially, the GE could describe all these phase transitions, and with critical exponents more general than those in mean field theory.

\par
\r{A}man et al. \cite{Aman2015,Aman2006} employed the scaled form in $d$ dimensions:

\begin{equation} \lambda^{d-2} S(M,J,Q)=S( \lambda^{d-3} M, \lambda^{d-2} J, \lambda^{d-3} Q),\label{180} \end{equation}

\noindent where $M$ is the mass, $J$ is the angular momentum, and $Q$ is the electric charge. $M$ is positive, but $J$ and $Q$ may have either sign. This scaling is satisfied by, for example, the Kerr-Newman black hole, an example that demonstrates nicely the unconventional BHT size scaling. However, the scaling in this paper is more general than that in Eq. (\ref{180}), and it is not tied to any scaling of the system size.

\section{Canonical classical models}

In this section, I discuss two simple black hole solutions to the Einstein field equations: the Kerr and the RN black holes. I will refer to these solutions as the {\it canonical classical models} (CCM). Both models have GHF entropies.

\par
The CCM, though not quantum, might offer some guidance about what to expect in real black holes. Their thermodynamic properties follow from the three variable Kerr-Newman black hole with \cite{Davies1977, Smarr1973}:

\begin{equation} S(M,J,Q)=\frac{1}{8} \left(2 M^2-Q^2+2 \sqrt{M^4-J^2-M^2 Q^2}\right).\label{190}\end{equation}

\noindent I work in length units, with the conversion to real units described in, for example, ref. \cite{Ruppeiner2007}.

\subsection{Kerr black hole}

The Kerr black hole is perhaps the best model for real astrophysical black holes. It has $Q=0$, for which Eq. (\ref{190}) gives

\begin{equation}S(M, J)=\frac{1}{4} M^2 \left(1+\sqrt{1-z^2}\right),\label{200}\end{equation}

\noindent where

\begin{equation}z=\frac{J}{M^2}.\label{210}\end{equation}

\begin{center}
\begin{figure}
\begin{minipage}[b]{0.5\linewidth}
\includegraphics[width=2.9 in]{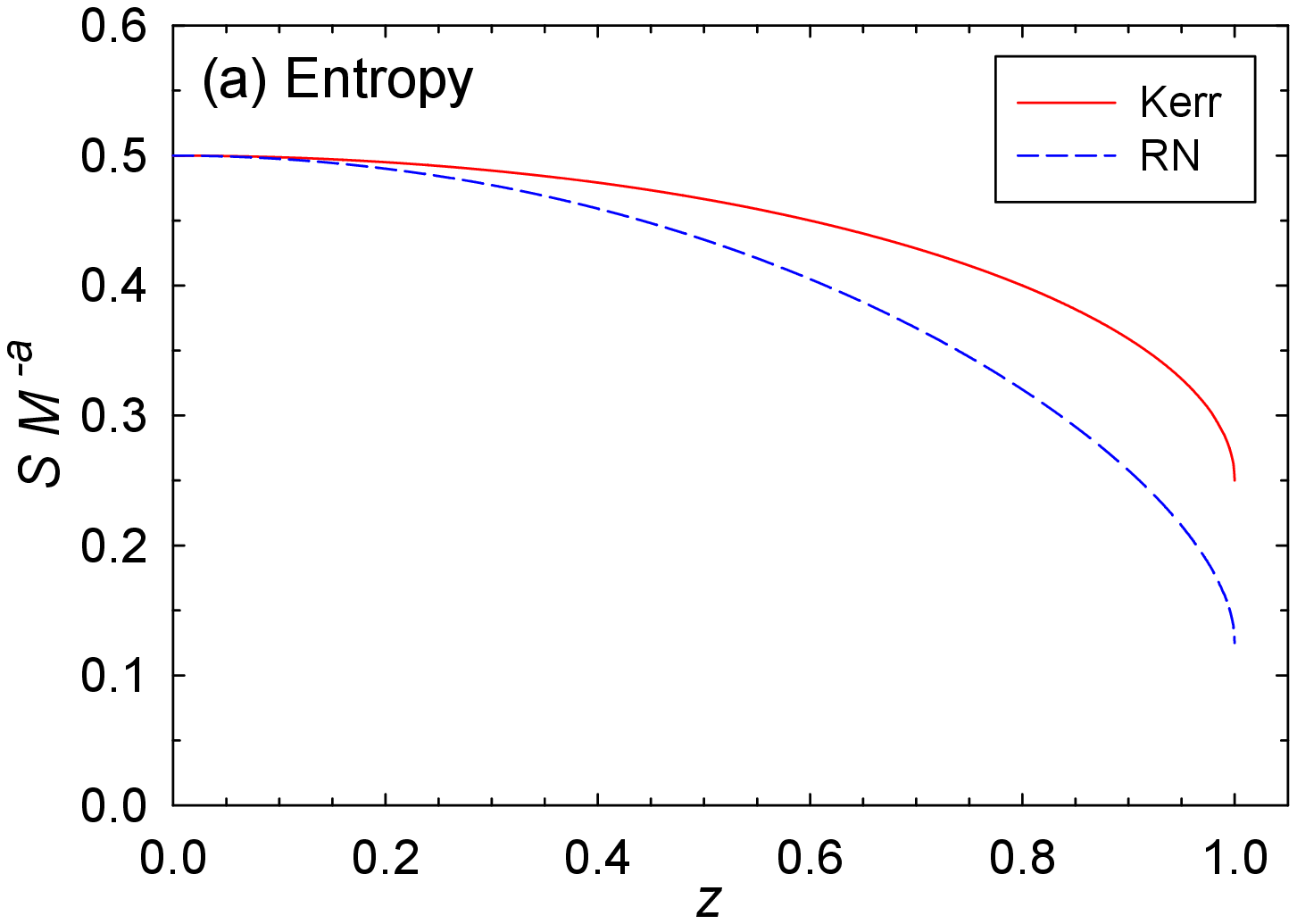}
\end{minipage}
\hspace{0.2 in}
\begin{minipage}[b]{0.5\linewidth}
\includegraphics[width=2.9 in]{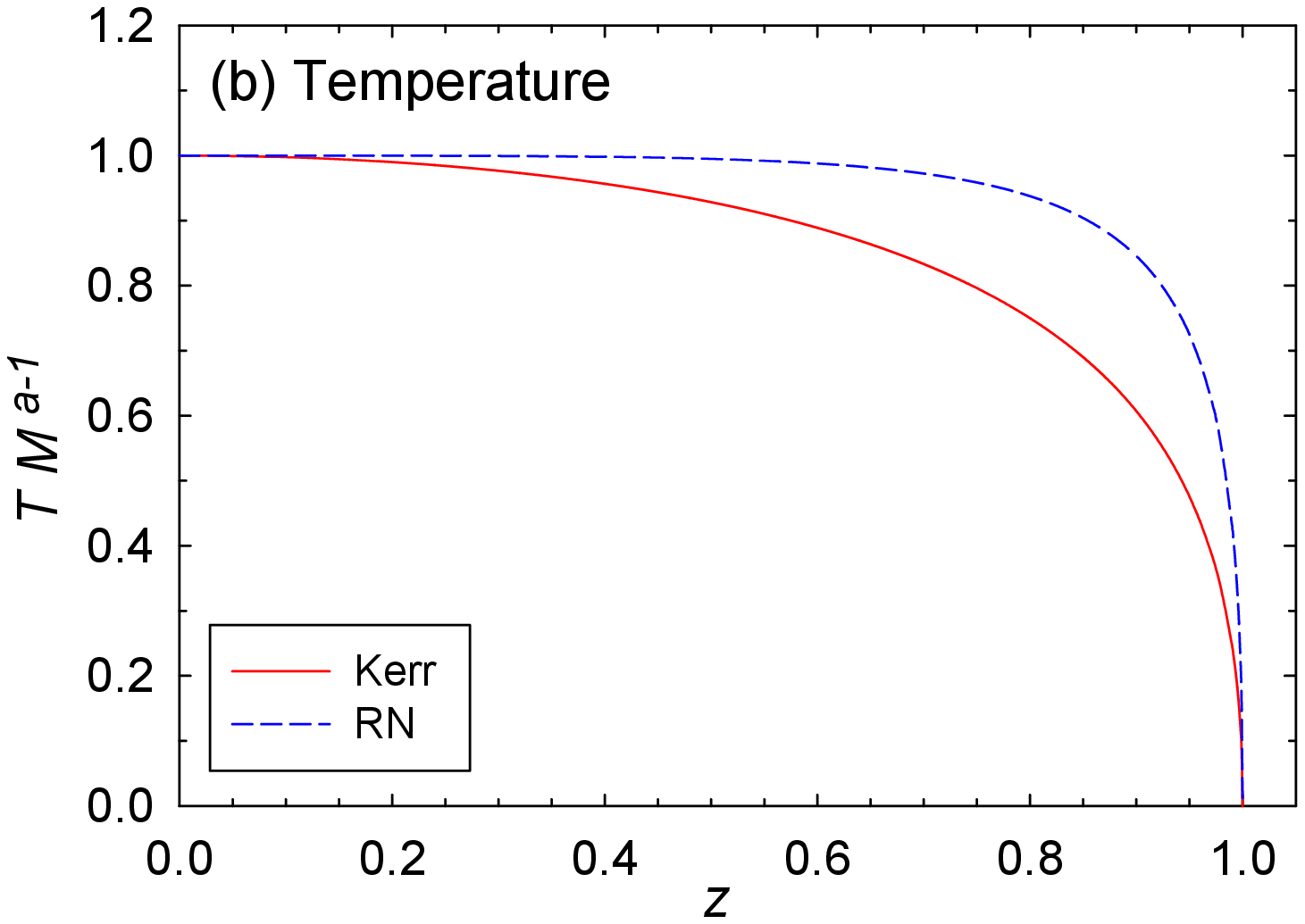}
\end{minipage}
\vspace{-0.05 in}
\begin{minipage}[b]{0.5\linewidth}
\includegraphics[width=2.9 in]{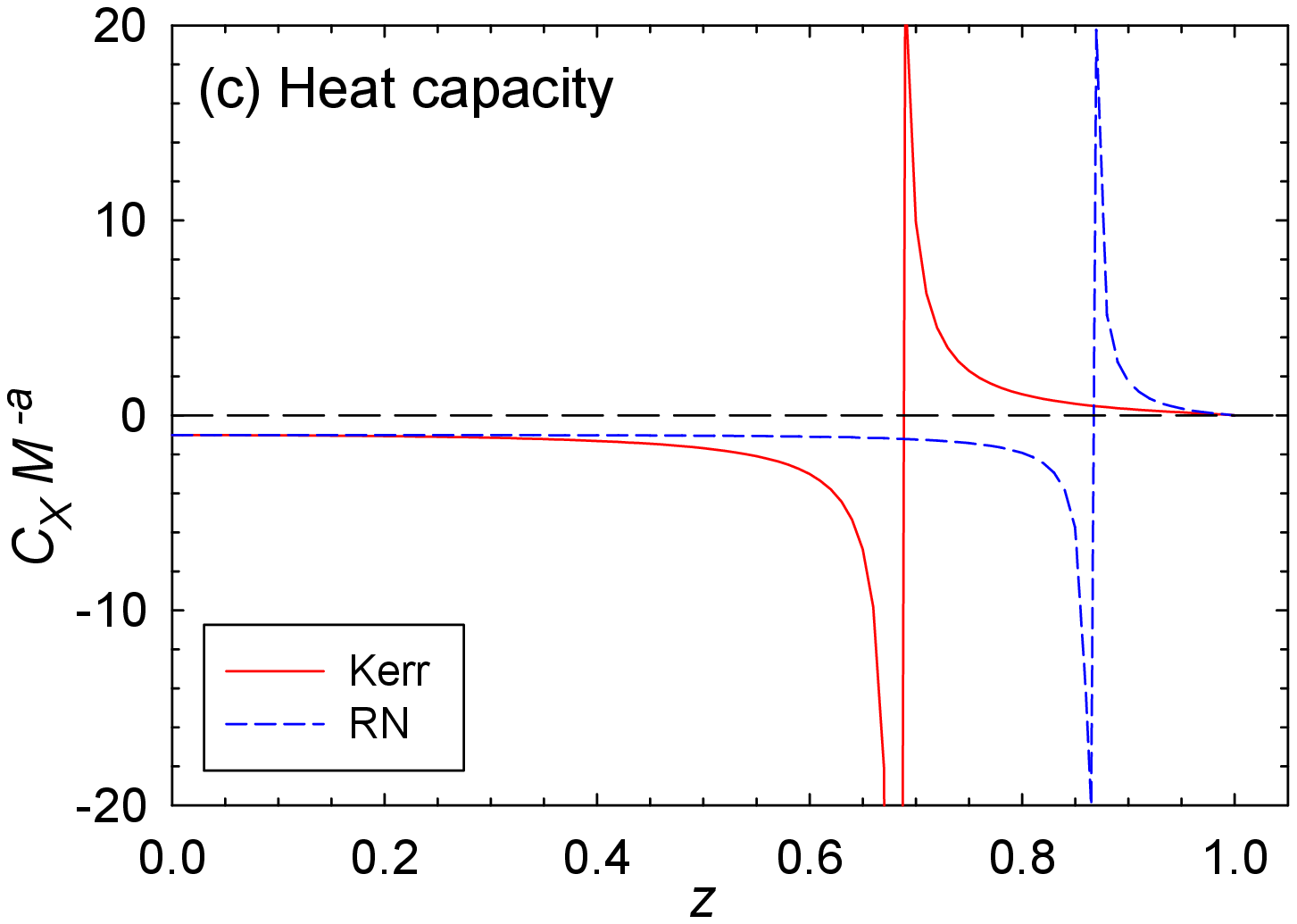}
\end{minipage}
\hspace{0.2 in}
\begin{minipage}[b]{0.5\linewidth}
\includegraphics[width=2.9 in]{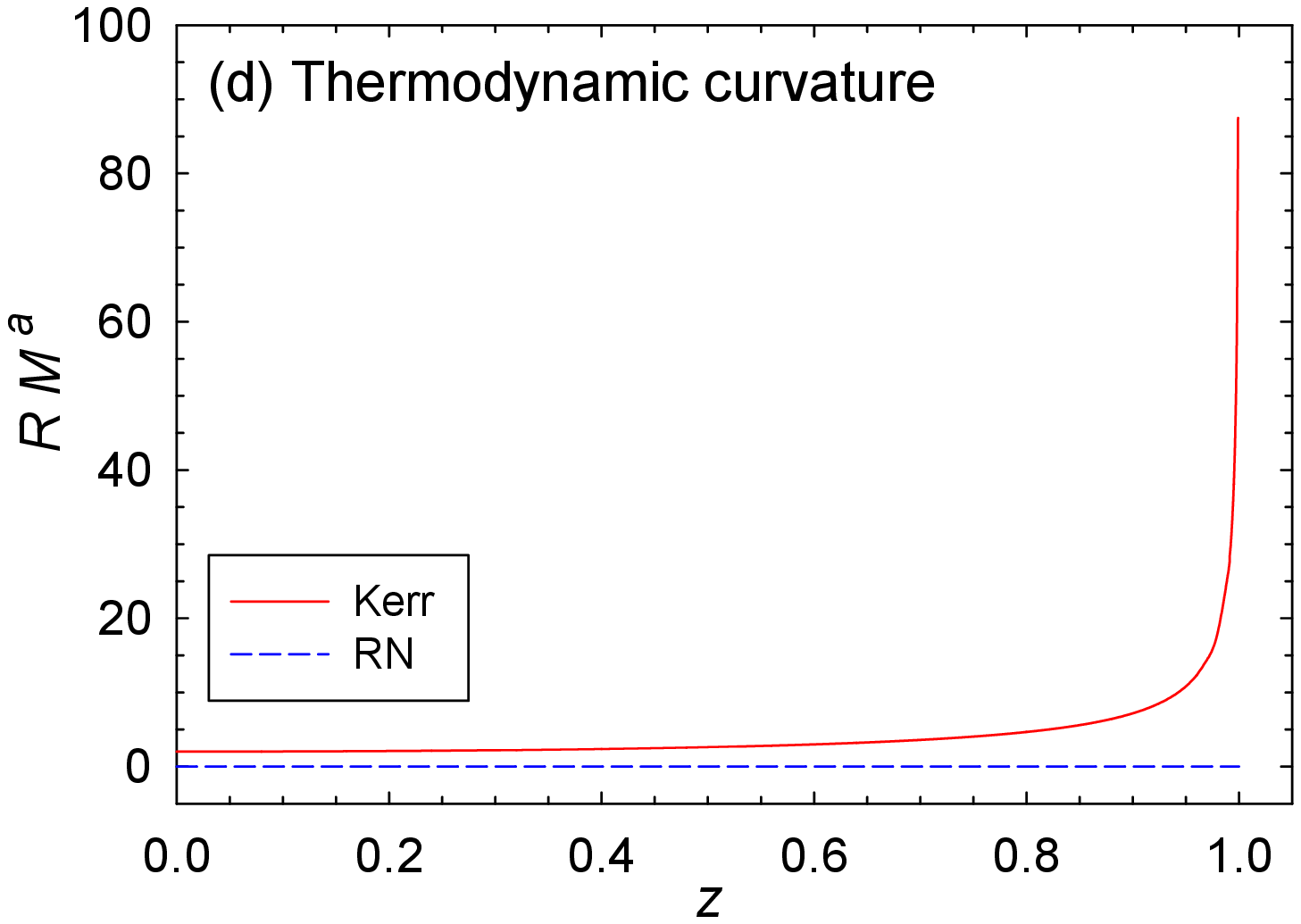}
\end{minipage}
\vspace{-0.05 in}
\begin{minipage}[b]{0.5\linewidth}
\includegraphics[width=2.8 in]{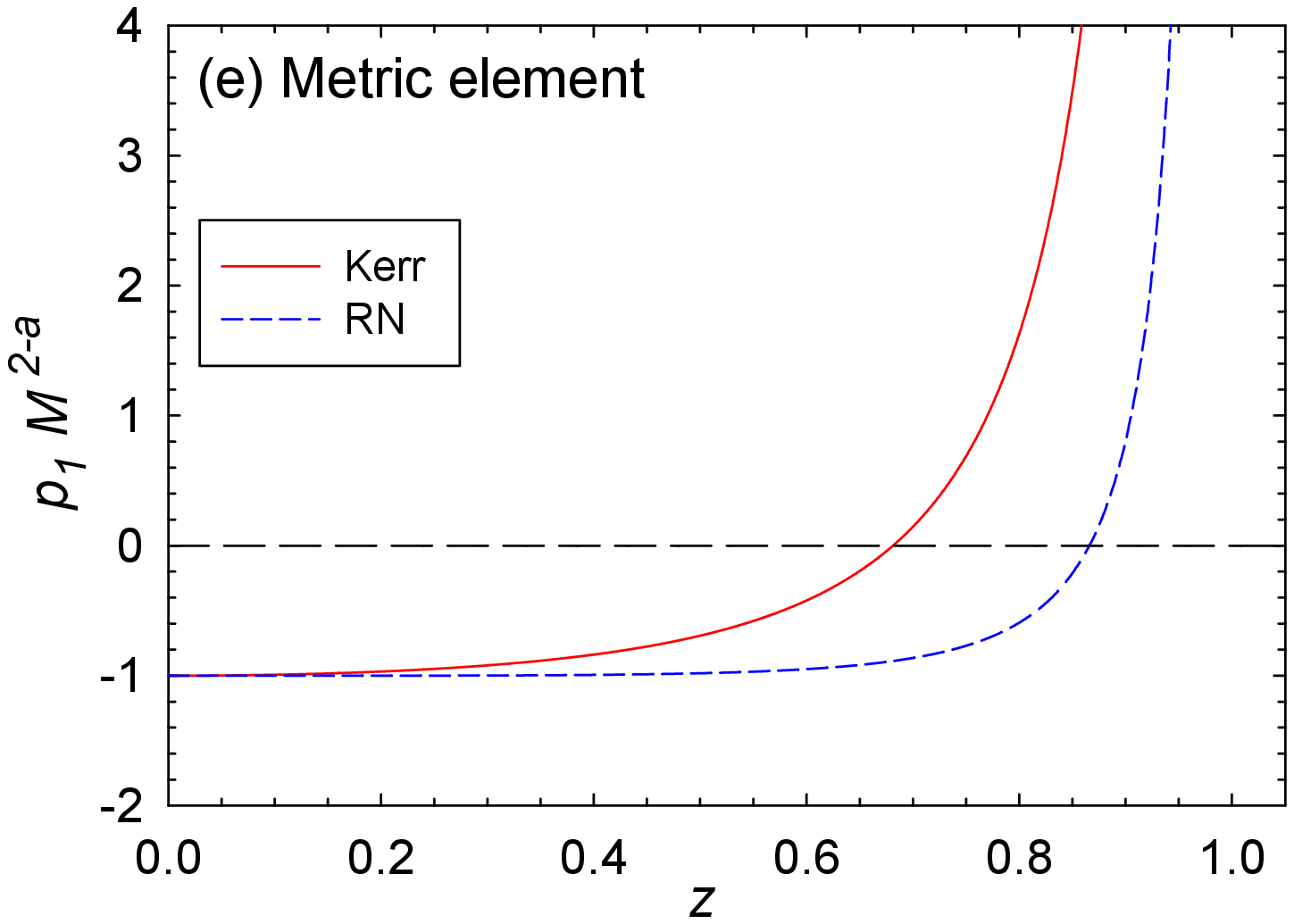}
\end{minipage}
\hspace{0.2 in}
\begin{minipage}[b]{0.5\linewidth}
\includegraphics[width=2.8in]{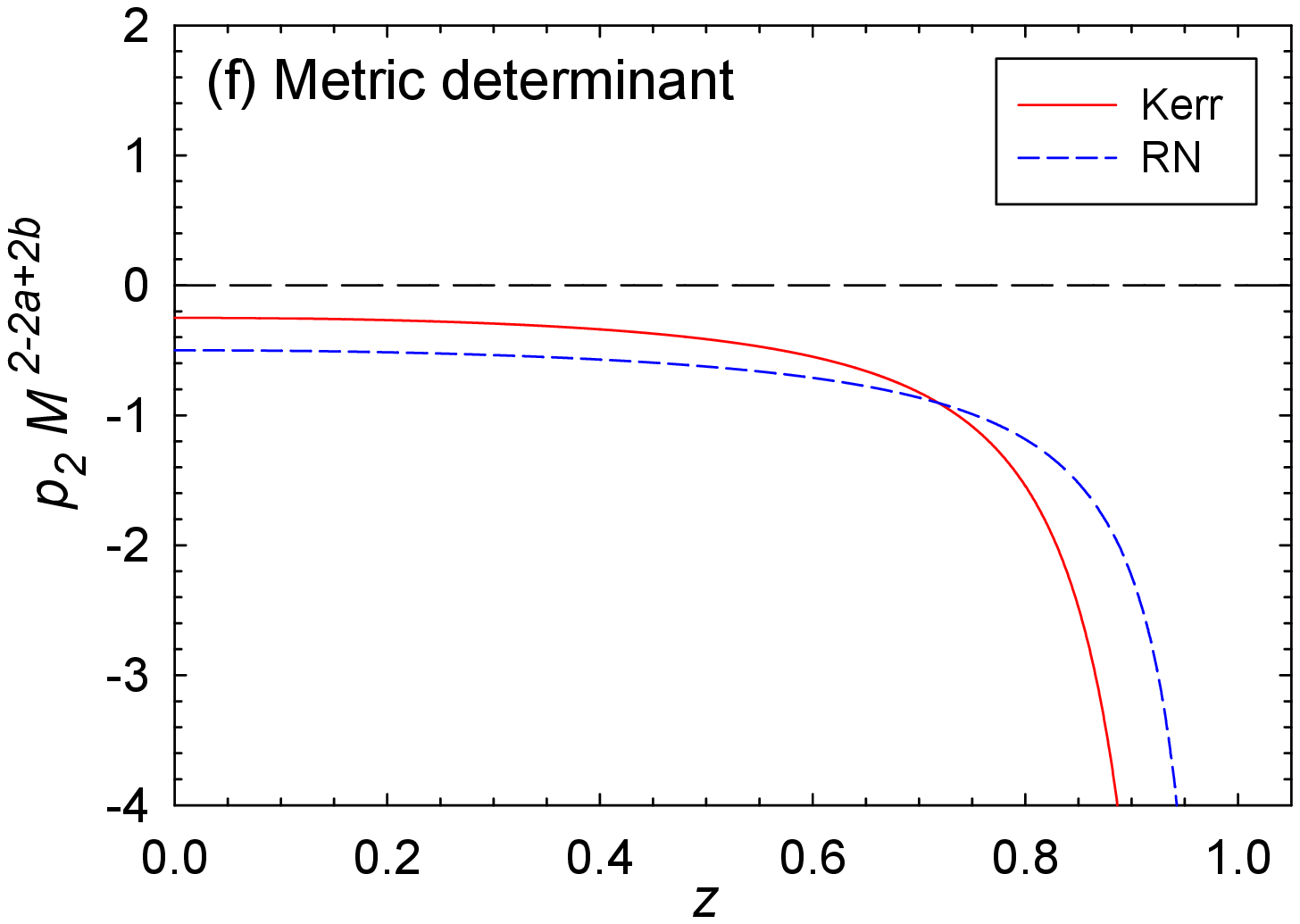}
\end{minipage}

\caption{Six functions evaluated for Kerr and RN: (a) the entropy $S$; (b) the temperature $T$; (c) the heat capacity $C_X$, with $X=J$ or $Q$; (d) the thermodynamic curvature $R$; (e) the metric element $p_1$; and (f) the metric determinant $p_2$. With the exception of $R$, Kerr and RN show qualitatively similar behavior.}
\label{fig:Figure3}
\end{figure}
\end{center}

\noindent Clearly, the exponents in Eq. (\ref{160}) are $(a,b)=(2,2)$. Figure \ref{fig:Figure3} shows six BHT functions for Kerr. These functions are all symmetric about $z=0$, with $-1<z<1$, and with $z=\pm 1$ corresponding to extremality $T=0$.

\subsection{Reissner-Nordstr$\ddot{\mbox{o}}$m black hole}

The RN black hole has $J=0$, for which Eq. (\ref{190}) gives

\begin{equation} S(M,Q)=\frac{1}{8}M^2 \left(2-z^2+2 \sqrt{1-z^2}\right),\label{220}\end{equation}

\noindent where

\begin{equation} z = \frac{Q}{M}.\label{230}\end{equation}

\noindent Clearly, the exponents in Eq. (\ref{160}) are $(a,b)=(2,1)$. Figure \ref{fig:Figure3} shows six functions for RN. As with Kerr, these functions are symmetric about $z=0$, with $-1<z<1$, and with $z=\pm 1$ corresponding to extremality. For RN, we have directly from Eq. (\ref{120}) that $R=0$. Indeed, \r{A}man et al. \cite{Aman2006} placed scaled forms of the entropy with $b=1$ into a special category having flat geometries.

\subsection{Some general observations}

\par
The CCM lack quantum properties, and thus likely fall short of a true physical representation. One might think that for sufficiently large black holes the effects of quantum mechanics will disappear, and the general relativity models will be correct. But such wishful thinking is not correct even in ordinary thermodynamics; for example, the quantum superconductivity and superfluidity persist all the way up to macroscopic length scales.

\par
Nevertheless, we might hope that the CCM offer at least some guidance for understanding real scenarios. Since the thermodynamic approach described in this paper offers many possible solutions, most of them probably not physical, the choices must be narrowed. My try is to generate solutions over a grid of reasonable $(a,b)$ values, and then to focus on the solutions that approximate most closely our expected black hole properties. The hope is that the true black hole behavior will be among these solutions. My discussion is limited to two independent variables, and to certain assumptions about the symmetry and analyticity of the solutions, but other choices are certainly possible.

\par
Experimental black hole data would be of great value in narrowing down the choices of possible solutions. Data is emerging from measurements of gravitational waves \cite{Abbott2016} and from tidally distorted events \cite{Gezari2012}. In addition, there are a host of good string theory models on the scene. The structure outlined in this paper could form a context for analyzing information from various sources.

\par
Here are a number of general observations about the CCM. Henceforth, I use the notation $M=X_1$ and $X=X_2$.

\noindent 1. $S$ is an even GHF of $z=X M^{-b}$, and analytic at $z=0$. See Eqs. (\ref{200}) and (\ref{220}). 

\noindent 2. The CCM have sharp extremal edges. See Figures \ref{fig:Figure3}(a) and \ref{fig:Figure3}(b). As $z$ gets larger, $S M^{-a}$ decreases from maxima at $z=0$ to finite positive values at $z=1$ at which $T\to 0$. This point marks the extremal point, where the surface gravity is expected to be zero \cite{Wald2001}.

\noindent 3. As $M\to 0$, $T\to\infty$ only if $a>1$. See Eq. (\ref{440}). This limit is Hawking's well-known result.

\noindent 4. The heat capacities $C_X$ start negative. See Fig. \ref{fig:Figure3}(c). This negative start reflects the well-known lack of thermodynamic stability of spherical self-gravitating objects.

\noindent 5. The heat capacities $C_X$ have infinities unconnected with the extremal limit at a single value of $|z|<1$. See Fig. \ref{fig:Figure3}(c). These infinities correspond to a change in sign of $C_X$ from negative to positive as $z$ increases. The heat capacities go to zero in the extremal limit $|z|=1$.

\noindent 6. The Kerr extremal black hole has thermodynamic curvature $R\to+\infty$. See Fig. \ref{fig:Figure3}(d). In ordinary thermodynamics, the Bose/Fermi ideal quantum gases have $R$ diverging to minus/plus infinity as $T\to 0$, and the effects of the quantum statistical interactions become more pronounced \cite{Mrugala1990, Oshima1999}. $R\to\infty$ as $T\to 0$ indicates that the conjectured fundamental particles constituting Kerr might be fermionic \cite{Ruppeiner2008}.

\noindent 7. The RN black hole has $R = 0$. See Fig. \ref{fig:Figure3}(d). This result suggests a classical non-quantum ideal gas as constituting the microscopic constituents. Such an idea may strike one as demonstrably unphysical \cite{Quevedo2008} since we associate black holes with strong gravitational interactions. But the gravitational component might be non-statistical, associated perhaps with the massive, central singularity predicted by general relativity.

\noindent 8. There are no regimes of thermodynamic stability. See Figs. \ref{fig:Figure3}(e) and \ref{fig:Figure3}(f). In particular, $p_2$ is always negative.

\subsection{Additional comments}

The sign change, and the corresponding plus and minus infinities, in the heat capacity $C_X$ have been interpreted as indicating phase transitions by Davies \cite{Davies1977}. But this interpretation has been questioned, since this sign change does not correspond to a change in thermodynamic stability \cite{Aman2003}. Nevertheless, the sign change in $C_X$ is a persistent feature in BHT, and I will come back to it below.

\par
There are certainly BHT models with true, non-extremal, phase transitions. Examples are the RN-AdS black hole, with a van der Waals type phase transition, and matter-free AdS space with a Hawking-Page phase transition between black hole/no black hole solutions. Results in the literature have $R$ typically negative at the der Waals type BHT phase transitions \cite{Aman2003, Shen2007, Sahay2010b, Niu2012}, the same sign as the critical phenomena models in ordinary thermodynamics. The Hawking-Page phase transition has been associated with $R=0$ \cite{Sahay2010b}. 

\par
Phase transitions other than extremal do not appear in my solutions below. But the method laid out here could be readily extended to include such cases.

\section{Calculation method}

In this section, I solve the GE in the context of the scaling relation Eq. (\ref{160}) that in the present notation is

\begin{equation} S=M^a Y\left(\frac{X}{M^b}\right).\label{240} \end{equation}

\noindent Eqs. (\ref{110}) and (\ref{120}) yield the thermodynamic curvature $R$, and Eqs. (\ref{140}) and (\ref{145}) yield the free energy $\phi$. Since $R$ has derivatives of $S$ up to third order in $Y(z)$, the resulting GE Eq. (\ref{130}) becomes a third-order ODE for $Y(z)$.

\par
Motivated by observation 1 in Section 3.3, I assume that $Y(z)$ is an analytic, even function at $z=0$, with series

\begin{equation} Y(z)=y_0+y_2 z^2 + y_4 z^4 +\cdots, \label{250} \end{equation}

\noindent where $y_0$, $y_2$, $y_4$, $\dots$ are the constant series coefficients. Substituting this series into $R$ and $\phi$ yields

\begin{equation} \displaystyle R M^a=-\frac{(b-1) (a-2 b)}{(a-1) a y_0}\,+\displaystyle\frac{\mathcal N_2}{(a-1)^2 a^2 y_0^2 y_2} z^2+O\left(z^4\right),\label{260}\end{equation}

\noindent where

\begin{multline}
\mathcal N_2=2 (b-1)\times
 \\
\!\!\!\!\!\!\!\!\!\!\!\! \left\{-a^3 \left(y_2^2-3 y_0 y_4\right)+a^2 \left([6 b-1] y_2^2-3 y_0 y_4\right)+4 a (1-3 b) b y_2^2+4 b^2 (2 b-1) y_2^2\right\},
\label{270}
\end{multline}

\noindent and

\begin{equation}\frac{M^a}{\phi}=-\frac{1}{(a-1) y_0}+\frac{y_2 (1+a-2 b)}{(a-1)^2 y_0^2} z^2+O\left(z^4\right).\label{280}\end{equation}

\par Substituting these two series into the GE Eq. (\ref{130}), and equating the zero'th order terms, yields

\begin{equation}\kappa=-\frac{(b-1)(a-2b)}{a}, \label{290}\end{equation}

\noindent independent of any series coefficients. Since the GE is a third-order ODE, three free constants are required for its integration. One of these constants has already been set to zero to make the series even. Specifying $y_0$ and $y_2$ now leads to a complete series solution, since equating the second-order series terms yields $y_4$, and equating higher-order terms yields all of the remaining coefficients. This series solution may be used to calculate a quadruple of values $\{z_0,Y(z_0),Y'(z_0),Y''(z_0)\}$ that serve as the initial conditions for a numerical solution of the GE for any values of $\{a,b,y_0,y_2\}$.

\par
Generally, we may easily prove that for any $(a,b)$, and for any nontrivial pair of constants $n_1$ and $n_2$, if $Y(z)$ is a solution, then $n_1 Y (n_2 z)$ is a solution \cite{Ruppeiner1998}. In critical phenomena theory such a result is called universality since it allows for a number of different solutions with the same critical exponents and the same function $Y(z)$. Whether or not this universality has physical significance in the BHT scenario is not yet clear. However, it simplifies the solutions process, since to find scaled solutions $Y(z)$ it suffices merely to give values for $(a,b)$ and to specify the signs of $y_0$ and $y_2$.

\par
I conclude this section by writing the full ODE for the exponent values $(a,b)=(2,2)$:

\begin{equation} Y^{(3)}(z)=\frac{z Y''(z)^2}{z Y'(z)-Y(z)}\,. \label{300}\end{equation}

\noindent To write the ODE for general $(a,b)$ takes too much space.\footnote{The denominator of Eq. (\ref{300}) is proportional to the inverse temperature given in Eq. (\ref{440}) for $(a,b)=(2,2)$. But this proportionality does not generalize to other values of $(a,b)$, so it is of little interest.}

\section{Results}

In this section, I discuss results. I confined my search of exponent values to the rectangle with diagonal corners at $\{a,b\}=\{1,1\}$ and $\{3,5/2\}$. The lower exponent limits come from the considerations discussed below, and the upper limits were set to avoid drifting too far from the CCM exponents.

\subsection{General comments}

\par
The GE admits a broad scope of possible solutions, reflecting the multiple physical possibilities for microscopic forces. However, in the presence of strong effective microscopic forces (i.e., those producing long-range correlations), the system arranges itself into spatially large mesoscopic structures of size $\xi$, in which the details of the microscopic forces are not important. This is the basic justification in the modern theory of critical phenomena for regarding only the dimensionalities of the physical space and of the order parameter to be relevant for determining the universal properties \cite{Stanley1999}. The result is a great simplification in the form of scaled universal expressions for the free energy. The information geometry of thermodynamic adds to this picture an explicit method of generating the scaled free energy in terms of the critical exponents.

\par
Since scaled expressions for the thermodynamic properties are also frequent occurrences in BHT (see Section 3), we might infer that the unknown underlying microscopic black hole constituents also interact strongly. In this event, the basic philosophy of critical phenomena holds for BHT, and the program for generating scaled solutions offered by solving the GE might prove effective. The expression in Eq. (\ref{240}), inspired by the CCM, seems the simplest try for a scaled solution.

\par
As I argued in Section 4, the only free parameters in the scaled solutions of the GE Eq. (\ref{130}), other than the simple scaling factors $n_1$ and $n_2$ in Section 4 (simply related to $y_0$ and $y_2$), are $(a, b)$, and the signs of $y_0$ and $y_2$. Guided by the considerations of Section 3, I will attempt a narrowing of the acceptable values of $(a,b)$.

\subsection{The signs of $y_0$ and $y_2$}

\par
Start by considering the signs of $y_0$ and $y_2$. First, notice from the series for $T$ Eq. (\ref{450}) in the Appendix that Hawking's result $T\to\infty$ as $M\to 0$ requires $a>1$. Include the series for $S$ Eq. (\ref{430}), and we see that positive $S$ and $T$ at $z=0$ requires positive $y_0$. Positive $y_0$ and $a>1$ are thus required. From the series Eq. (\ref{490}), $C_X$ will now be negative at $z=0$, a feature also of the CCM.

\par
The sign of $y_2$ is equally clear. The series for $S$ has $S M^{-a}$ be a local maximum at $z=0$ if and only if $y_2$ is negative. Such local maximas in $S$ obtain in the CCM. Furthermore, detailed numerical solutions to the GE for the $(a,b)$ values considered here show that $S$ is monotonically decreasing with increasing $z$ if and only if $y_2$ is negative. I thus consider only negative $y_2$.

\par
For given $(a,b)$, $Y(z)$ for any $(y_0, y_2)$ may be scaled to $Y(z)$ for any other $(y_0, y_2)$ with the same signs using the scaling relation Eq. (\ref{150}). Therefore, with no loss of generality, I take $y_0=1$ and $y_2=-1$ in all of my solutions below.

\subsection{The solution types}

\begin{figure}
\begin{minipage}[b]{0.5\linewidth}
\includegraphics[width=2.7in]{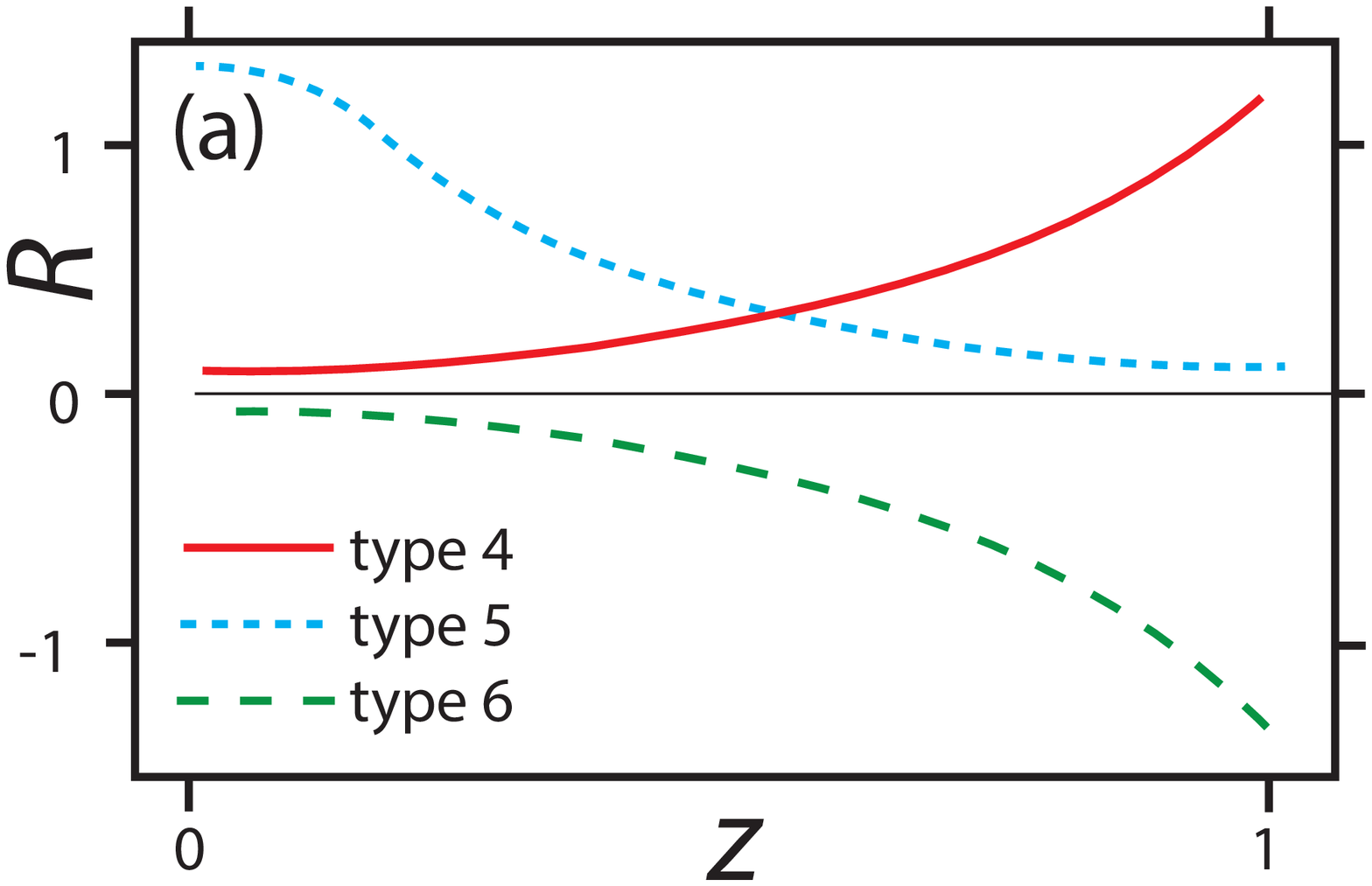}
\end{minipage}
\hspace{0.0 cm}
\begin{minipage}[b]{0.5\linewidth}
\includegraphics[width=2.7in]{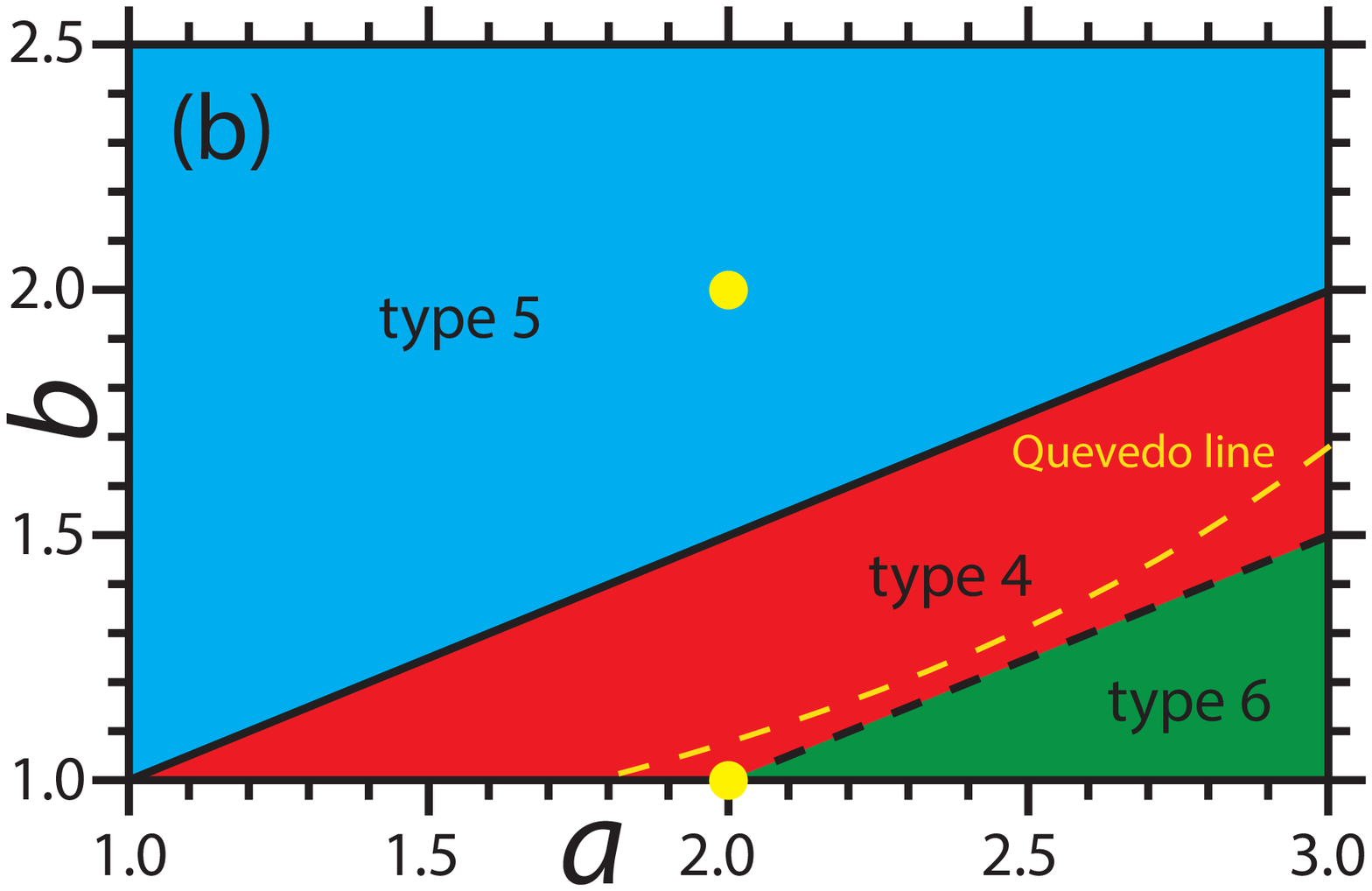}
\end{minipage}
\caption{Types of functions for $R$ sketched as $z$ increases from zero: (a) type $4$ (red) has $R$ diverging to +$\infty$ at some finite $z$; type $5$ (blue) has $R$ going asymptotically to zero; and type $6$ (green) has $R$ diverging to -$\infty$ at some finite $z$. (b) The regimes in $(a,b)$ space for which each type obtains. Along the solid line, $R$ is constant but not zero (type $3$), and along the dotted line, $R=0$ (type $2$). The axis $b=1$ is type $1$. The bold yellow dots correspond to the $(a,b)$ values of Kerr $(2,2)$ and RN $(2,1)$. The dashed yellow curve is the Quevedo line where $C_X$ and $R$ diverge at the same $z$. Type $4$ seems physically the most relevant.}
\label{fig:Figure4}
\end{figure}

The scaled solutions to the GE fall into basic types according to the values of $(a,b)$. I list these types in this subsection. Types $4-6$ seem the most physically relevant. Figure \ref{fig:Figure4} sketches the types, and the corresponding regimes of $(a,b)$ space where they obtain.

\par
Type $1$ has $b=1$, corresponding to $\kappa=0$, by Eq. (\ref{290}). In this case, the GE clearly has $R=0$. Furthermore, $R=0$ no matter what the function $Y(z)$. The case $b=1$ is a special category identified by \r{A}man et al. \cite{Aman2006} as having flat geometry. Since the GE tells us nothing about $Y(z)$ for $b=1$, I will not consider type $1$ further.

\par
Type $2$ has $a = 2b$, also corresponding to $\kappa=0$, and thus to $R=0$. One can show, however, that type $2$ does not allow just any $Y(z)$, and that the even solution to $R=0$ is

\begin{equation} Y(z) = y_0 + y_2 z^2, \label{310}\end{equation}

\noindent with $y_0$ and $y_2$ free constants. This solution yields

\begin{equation} T=\frac{M^{1-2b}}{2 b y_0}, \label{320}\end{equation}

\noindent independent of $z$. A $T$ independent of $z$ is quite unlike the $T's$ of the CCM in Fig. \ref{fig:Figure3}(b). Type $2$ is then perhaps not very physically relevant.

\par
Type $2$a has $(a,b)=(2,1)$ (the RN exponents), and so it is both type $1$ and type $2$. However, {\it any} function $Y(z)$ has $R=0$ for all three cases: $(a,b)=(2,1)$, $(a,b)=$ Limit$[(a,1)\to(2,1)]$, and $(a,b)=$ Limit$[(2,b)\to (2,1)]$. Hence, this hybrid case actually corresponds to type $1$.

\par
Type $3$ has $a=2b-1$. The simple quadratic expression for $Y(z)$ in Eq. (\ref{310}) yields

\begin{equation} R=\frac{M^{1-2b}}{2(2b-1) y_0},\label{330}\end{equation}

\noindent a constant independent of $z$. It also yields

\begin{equation}\phi=-2M^{2b-1} (b-1) y_0.\label{340}\end{equation}

\noindent Type $3$ thus has even solution $Y(z) = y_0 + y_2 z^2$, with $\kappa=(b-1)/(2b-1)$, a value for $\kappa$ consistent with Eq. (\ref{290}). Furthermore,

\begin{equation} S=M^{2b-1} (y_0 + y_2 z^2),\label{350}\end{equation}

\noindent and

\begin{equation} T=\frac{M^{2(1-b)}}{(2b-1) y_0 - y_2 z^2}. \label{360}\end{equation}

\noindent With $y_0>0$ and $y_2<0$, the expressions for $S$ and $T$ have features in common with the CCM, and thus type $3$ might have some physical interest.

\par
Turn now to the three other distinctive types $4-6$. For guidance, refer to the ODE in Eq. (\ref{300}), given for the case with $(a,b)=(2,2)$. Key here, and for more general values of $(a,b)$, is whether or not the denominator on the right-hand side crosses through zero. Unless the numerator is also zero at this crossing point, the result is a diverging $Y^{(3)}(z)$. Since $Y^{(3)}(z)$ is a factor in the numerator of $R$, but not in its denominator [see Eq. (\ref{540})], the zero crossing has $R$ also diverging. I identify three types of divergences: $R$ diverging positive (type $4$), negative (type $6$), or $R$ not diverging at all (type $5$). I denote by $z=z_R$ the position of a possible divergence of $R$.

\par
A diverging $R$ signals a nonanalytic solution for $Y(z)$. A number of examples with nonanalytic solutions were found in the ordinary thermodynamic examples shown in Table \ref{tab:table1}. There, they corresponded to phase transitions, and in some cases bridging techniques for continuing solutions beyond these phase transitions were found. However, there is too little guidance from the CCM to attempt such a bridging here, and I simply terminate solutions at points of diverging $R$. On the first try, it seems natural to associate these divergences with the extremal limit.

\par
Type $4$ has $(a+1)>2b>a$. $R$ starts positive at $z=0$, and then increases monotonically to $+\infty$ at some value $z=z_R$. This case matches qualitatively the situation in Kerr in Fig. \ref{fig:Figure3}(d), so type $4$ might be a serious candidate for real black holes.

\par
Type $5$ has $2b>(a+1)$. $R$ starts positive at $z=0$, and then goes asymptotically to zero with increasing $z$. Physically, $R$ going to zero asymptotically speaks to interparticle forces diminishing in the extremal limit. For type $5$, the solutions stay regular all the way out to $z\to\infty$. But we expect big spin and/or big charge to lead to a black hole breakup at some finite $z$. Hence, type $5$ might not be physically plausible.

\par
Type $6$ has $2b<a$. $R$ starts negative at $z=0$, and then decreases monotonically to $-\infty$ at some value $z=z_R$. Type $6$ does not match the CCM, so it is harder to fit it into a possible physical picture. Negatively diverging $R$'s speak to critical phenomenon type phase transitions, with effectively long-range attractive forces between microscopic constituents producing long-range order. Such phase transitions certainly take place in BHT models \cite{Chamblin1999, Sahay2010b}, but they are typically associated with more independent thermodynamic variables or with AdS backgrounds, neither present in this paper. Type $6$ also has $T$ increasing monotonically with $z$, quite unlike the decrease seen in the CCM. Hence, I do not regard type $6$ as physically plausible.

\par
Figure \ref{fig:Figure5}(e) shows $R$ for $(a,b)=(1.9,1.4)$ (type $4$). Figure \ref{fig:Figure5}(f) shows the values $z=z_R$ where $|R|$ diverges for types $4$ and $6$. I consider the regime with $z>z_R$ to be unphysical because $Y(z)$ ceases to be analytic at $z=z_R$. Again, type $5$ has $z_R\to\infty$.

\subsection{Further discussion of solution types}

\begin{figure}
\begin{minipage}[b]{0.5\linewidth}
\includegraphics[width=2.1in]{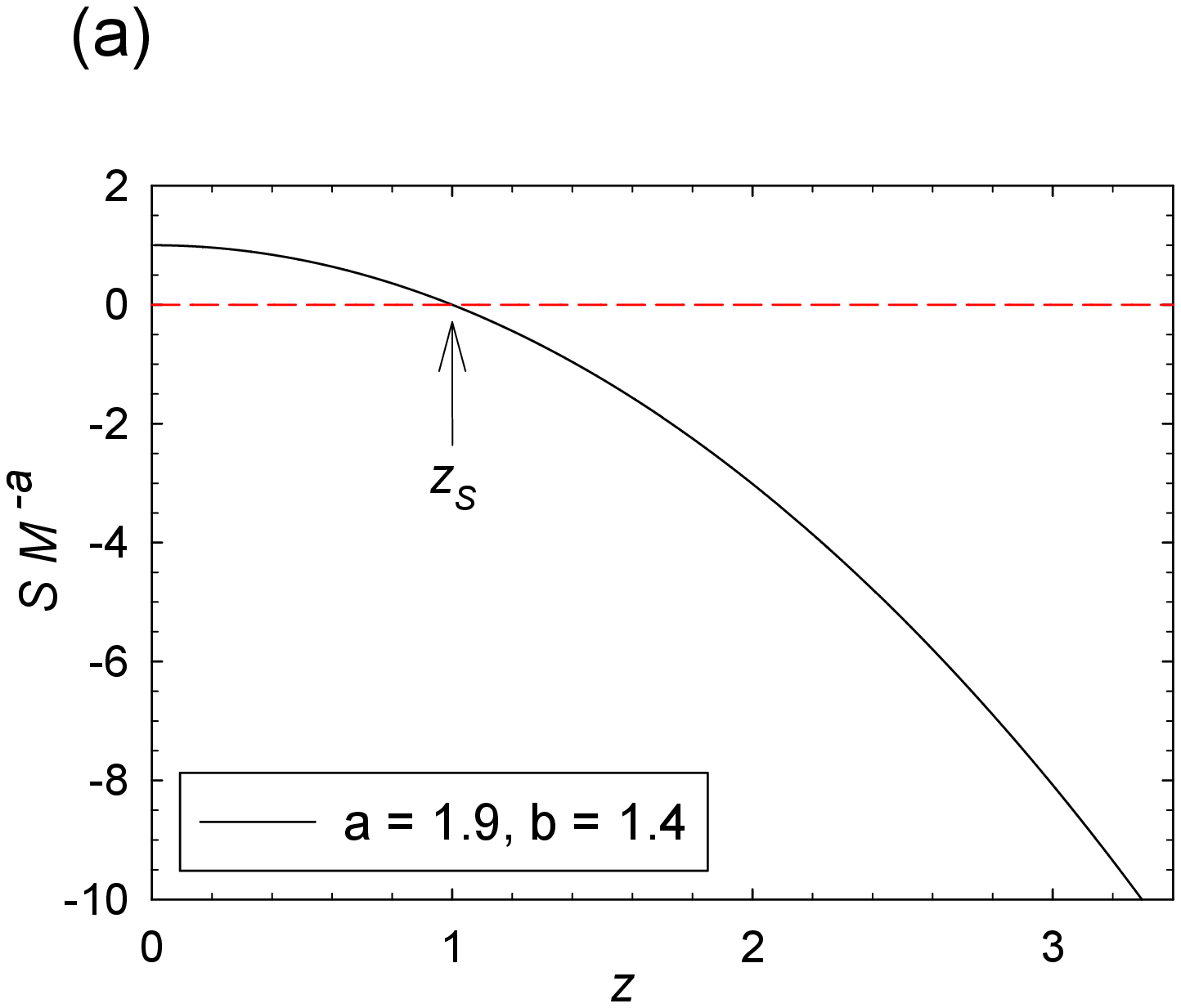}
\end{minipage}
\hspace{0.0 cm}
\begin{minipage}[b]{0.5\linewidth}
\includegraphics[width=2.1in]{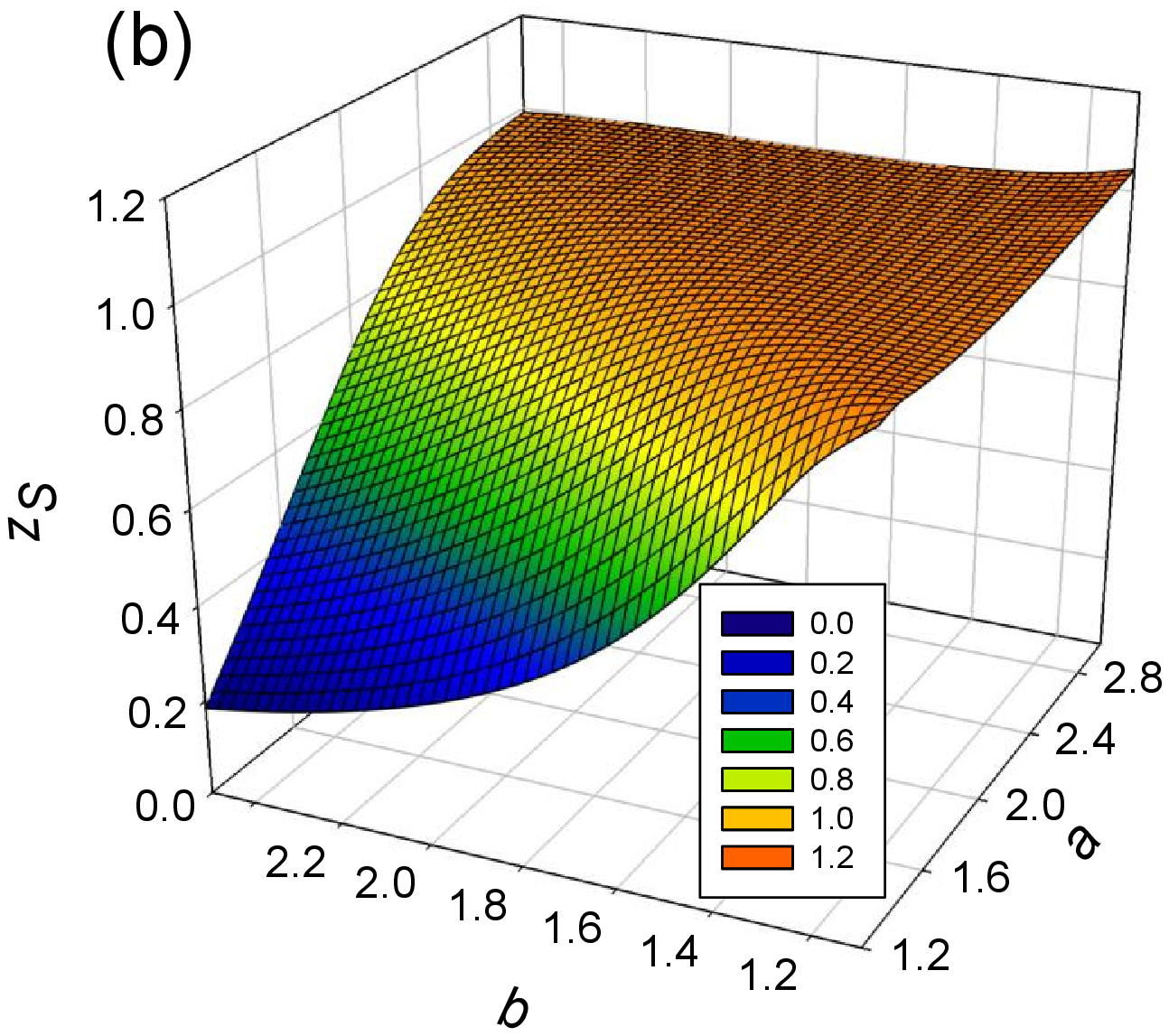}
\end{minipage}
\begin{minipage}[b]{0.5\linewidth}
\includegraphics[width=2.1in]{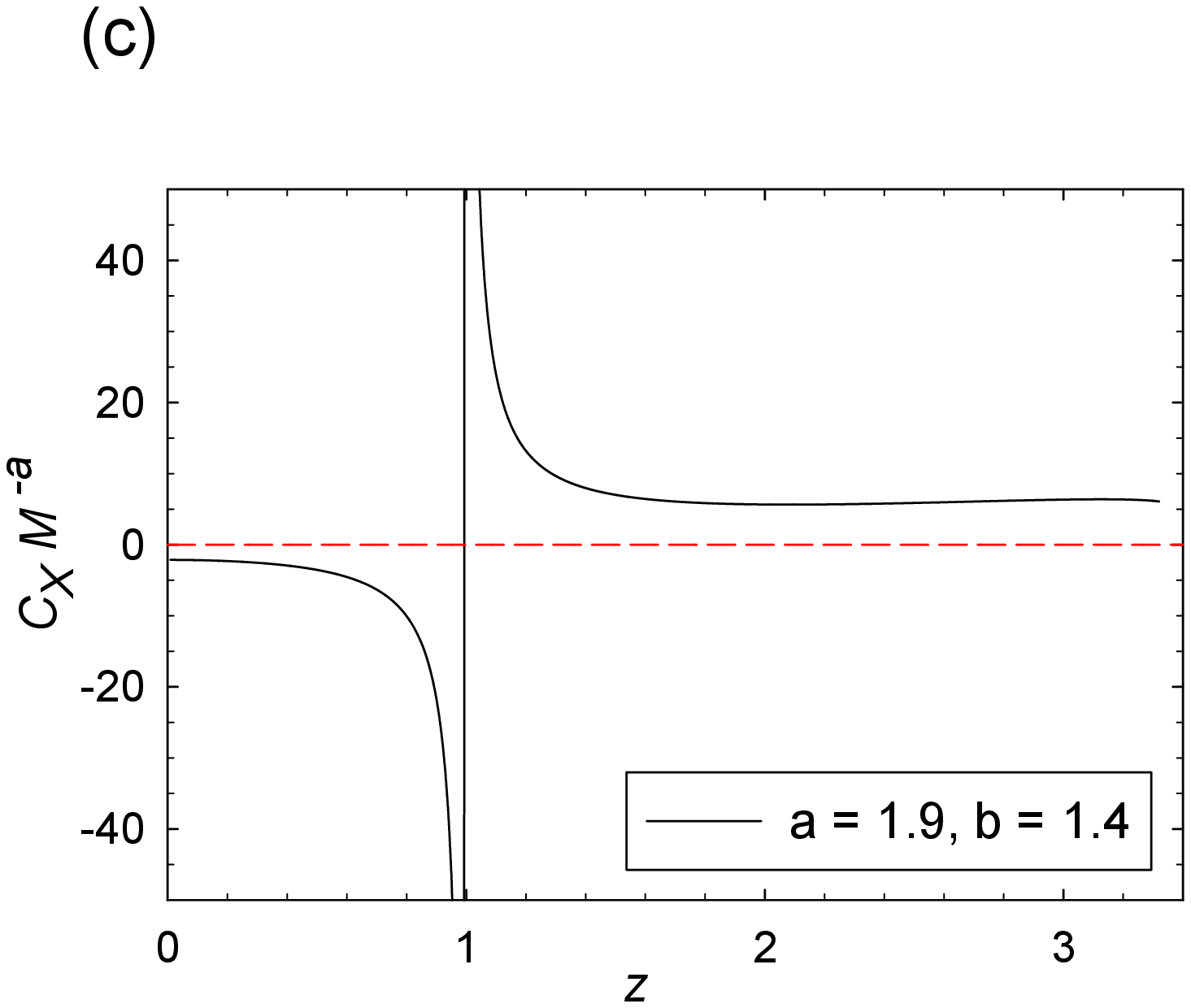}
\end{minipage}
\hspace{0.0 cm}
\begin{minipage}[b]{0.5\linewidth}
\includegraphics[width=2.1in]{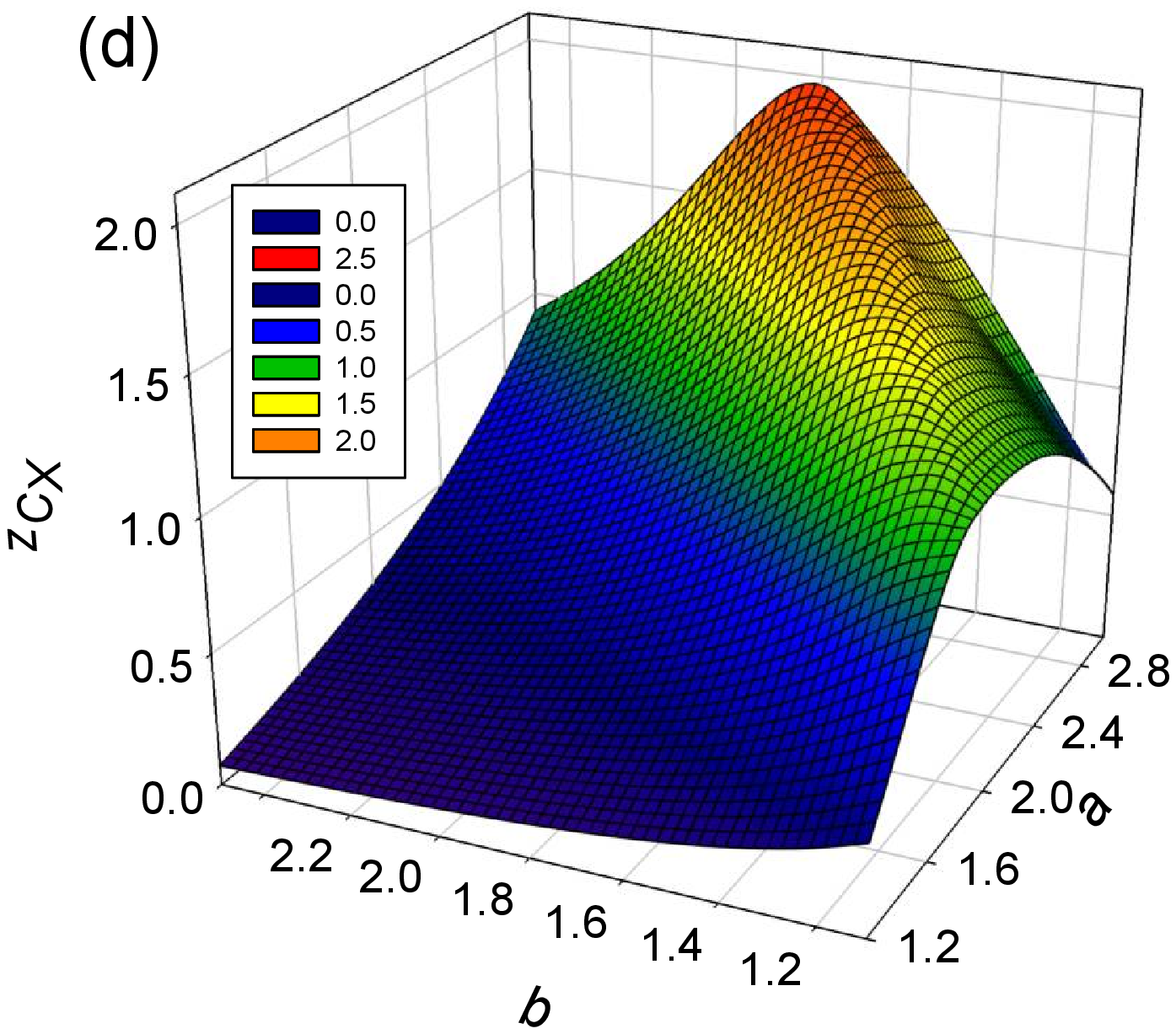}
\end{minipage}
\begin{minipage}[b]{0.5\linewidth}
\includegraphics[width=2.1in]{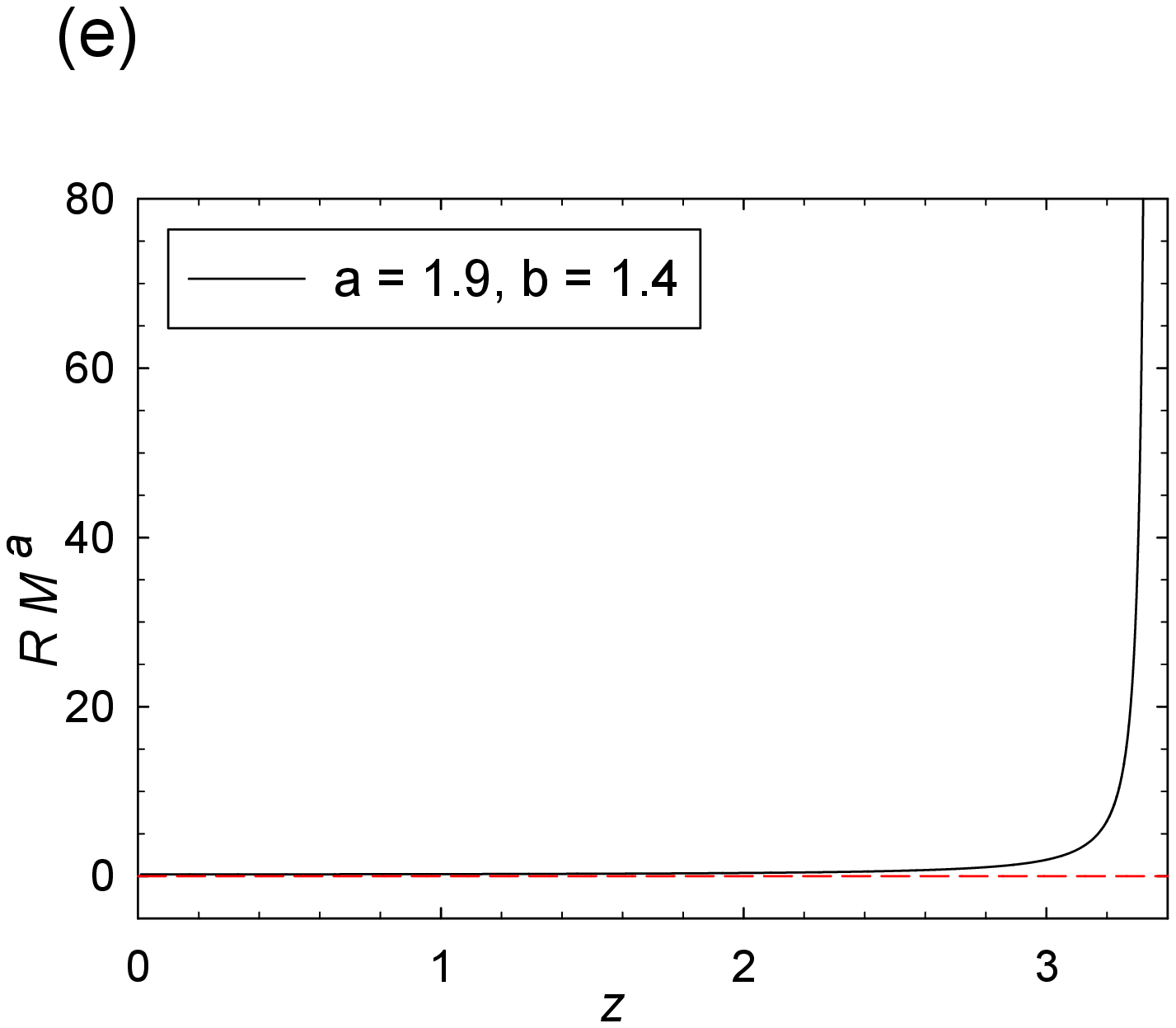}
\end{minipage}
\hspace{0.0 cm}
\begin{minipage}[b]{0.5\linewidth}
\includegraphics[width=2.1in]{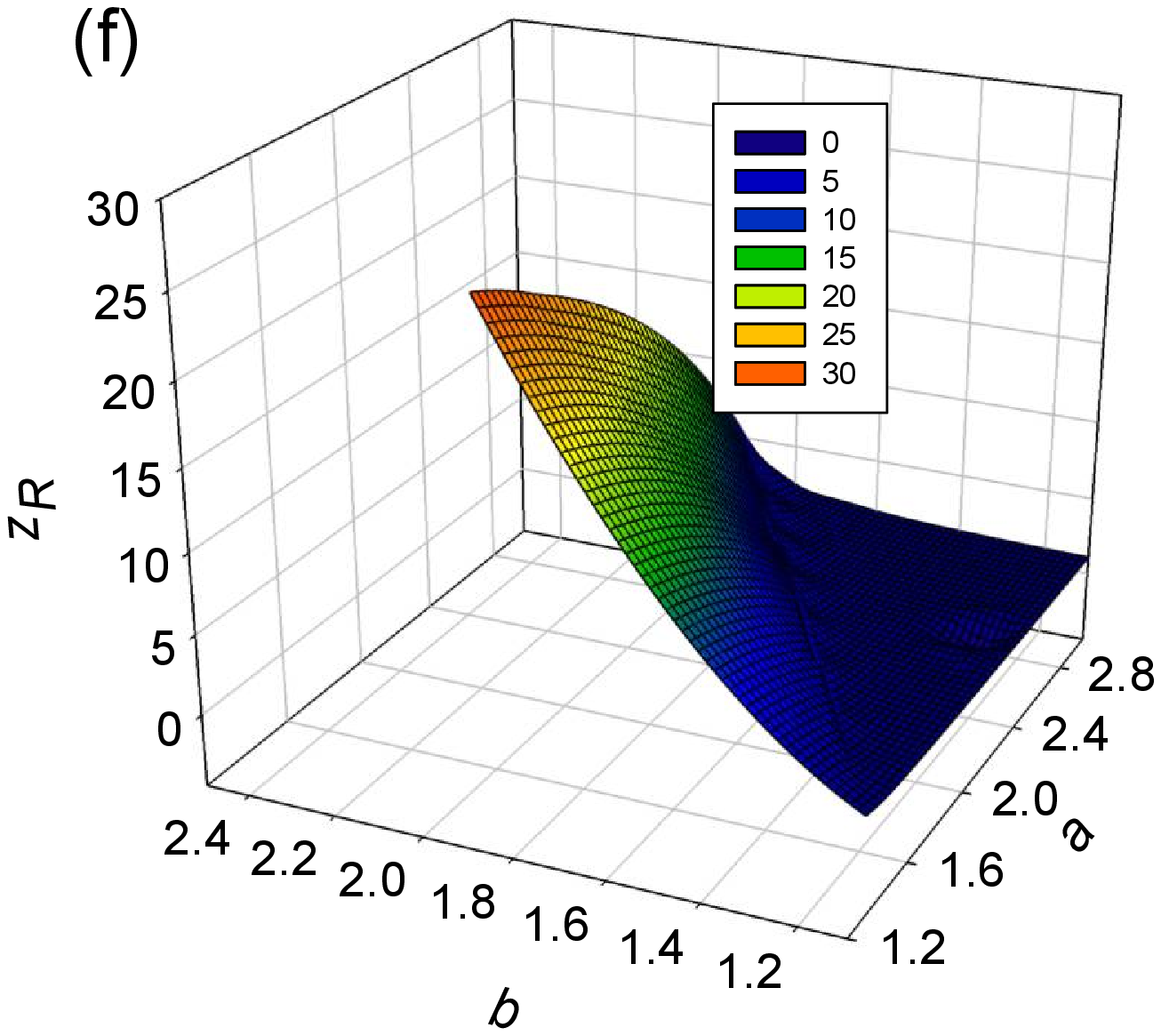}
\end{minipage}
\caption{Axis crossing $z_S$ and the divergence positions $z_{C_X}$ and $z_R$. Also shown are representative cases with $a=1.9$ and $b=1.4$, having $z_{C_X}=0.9970$ and $z_R=3.326$. Figures (a) and (b) represent the entropy $S$, with the $z$ axis crossings at $z=z_S$; (c) and (d) represent the heat capacity $C_X$, with the divergence positions at $z=z_{C_X}$ for the $(a,b)$ values corresponding to $z_{C_X}<z_R$; and (e) and (f) represent the thermodynamic curvature $R$, with the divergence positions at $z=z_R$ for types $4$ and 6. For type $5$, $z_R\to\infty$.}
\label{fig:Figure5}
\end{figure}

The simplest thermodynamic variable is $S$. For all of the values of $(a, b)$ considered here, $S=M^a Y(z)$ shows the qualitative behavior in Figure \ref{fig:Figure5}(a). For small $z$, the series Eq. (\ref{430}) has $Y(0)=1$, with $Y(z)$ decreasing monotonically with increasing $z$. Numerical solutions continue to show this monotonic decrease, with $Y(z)$ crossing the $z$ axis at some point $z=z_{S}$. Figure \ref{fig:Figure5}(b) shows $z_{S}$ as a function of $(a,b)$. All cases have $z_S<z_R$.

\par
The value $S=0$ could be taken to mark the absolute lower limit of the value of $S$, since negative entropies are hard to interpret physically. However, negative BHT entropies have been found as well in charged Gauss-Bonnet dS/AdS gravity by Odintsov et al. \cite{Nojiri2002, Nojiri2017}. These negative entropies were interpreted as marking either a regime where parameter values have transitioned into an unacceptable range, or into a regime where there is some sort of a phase transition. Clunan et al. \cite{Clunan2004} argued that negative entropies in Gauss-Bonnet gravity could be eliminated by adding to $S$ a constant positive factor. Such an addition certainly fits the spirit of the theory of critical phenomena in which only the ``singular parts'' of quantities are considered.

\par
The scaled behavior of $S$ in Fig. \ref{fig:Figure5}(a) shows some qualitative features in common with those of the CCM in Figure \ref{fig:Figure3}(a). However, $S$ for the CCM terminates at a value for $z$ with infinite slope, $Y'(z)\to -\infty$. This infinite slope leads to $T=0$, by Eq. (\ref{440}). $T=0$ corresponds to the extremal limit in the CCM, with $T$ proportional to the surface gravity.

\par
The extremal limit from the solutions to the GE is less well defined. The character of $T$ is indicated by the series Eq. (\ref{450}). Clearly, with $z$ increasing from $z=0$, $T$ decreases if $2b>a$ (corresponding to types $4$ and $5$ in Fig. \ref{fig:Figure4}), and increases if $2b<a$ (corresponding to type $6$). The full solution to the GE continues this monotonic decrease to larger $z$, including constant $T$ for $2b=a$. Clearly, type $6$ with its monotonically increasing $T$ is physically unsatisfactory.

\par
The final element of my picture is the divergence of the heat capacity $C_X$, with the case for the CCM shown in Fig. \ref{fig:Figure3}(c). Many GE solutions on my $(a,b)$ grid show such a divergence; see Figure \ref{fig:Figure5}(c) for a representative example. Neither for the CCM nor for the GE solutions does the divergence of $C_X$ typically correspond either to a divergence of $R$ or to a breakdown of analyticity in $Y(z)$. $C_X$ diverges only because its denominator passes through zero and changes sign. Nevertheless, the BHT literature pays considerable attention to the divergence of $C_X$, so I will do likewise.

\par
An interesting question is whether the divergence of $C_X$ at $z=z_{C_X}$ happens prior to that of $R$ at $z=z_R$. Both types $4$ and $6$ have finite $z_R$, and for the infinity in $C_X$ to be in the physical regime, as it is for the CCM, we must have $z_{C_X} < z_R$. Figure \ref{fig:Figure5}(d) shows $z_{C_X}$ as a function of $(a,b)$ for cases with $z_{C_X} < z_R$. Type $5$ is analytic all the way out to infinity, so for type $5$ we could have $z_{C_X}\to\infty$.

\par
The limiting curve in the $(a,b)$ plane with $z_{C_X}= z_R$ may be of interest to researchers who argue that a true BHT phase transition should have $C_X$ and $R$ diverging in the same place; see Quevedo \cite{Quevedo2008}. If this overlap of divergences is physically necessary, then the curve of overlap would logically correspond to the locus of true physical black holes. This curve is shown in Fig. \ref{fig:Figure4}, with points above the curve corresponding to $z_{C_X}<z_R$.

\par
\begin{figure}
\begin{minipage}[b]{0.5\linewidth}
\includegraphics[width=2.67in]{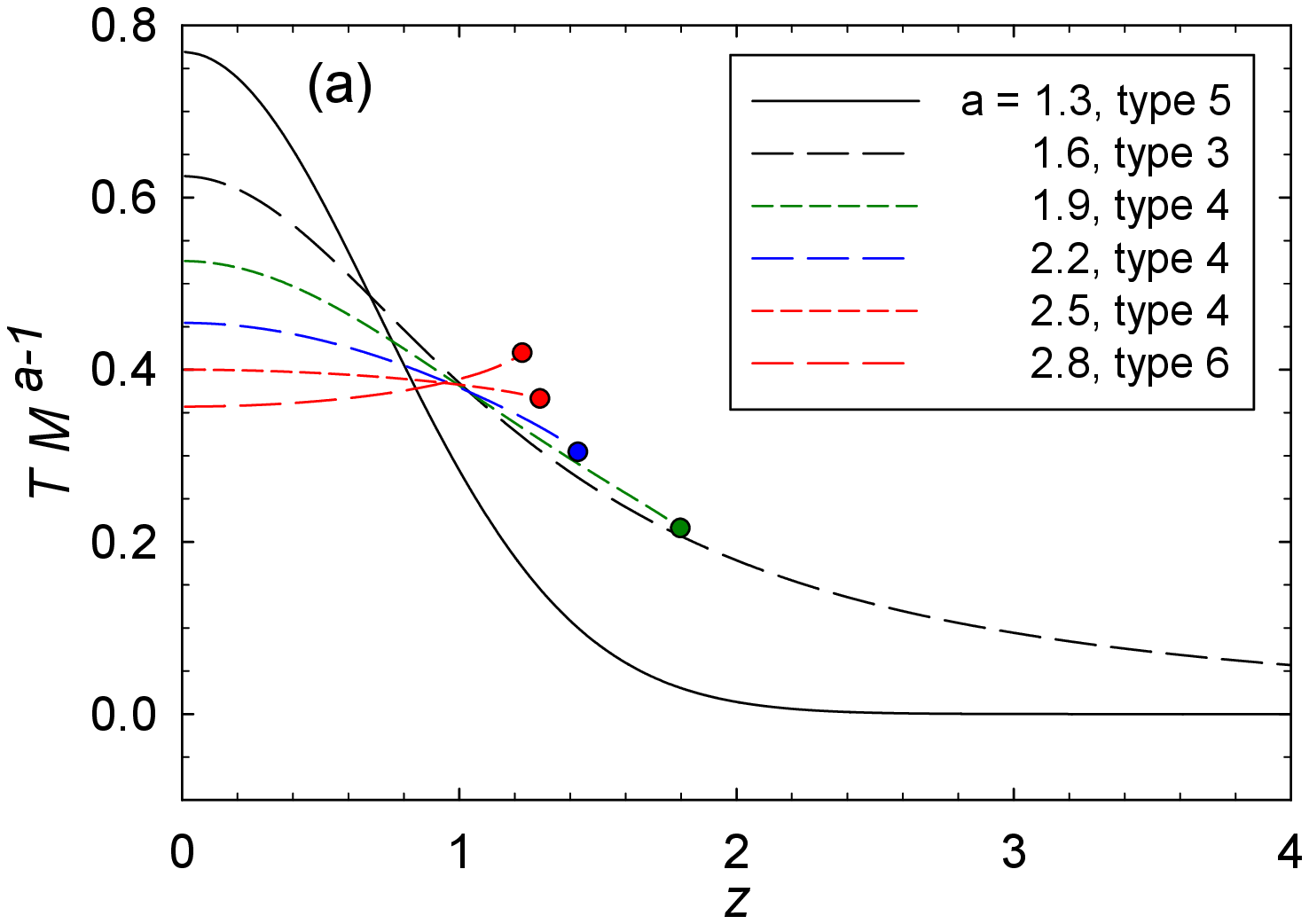}
\end{minipage}
\hspace{0.0 cm}
\begin{minipage}[b]{0.5\linewidth}
\includegraphics[width=2.7in]{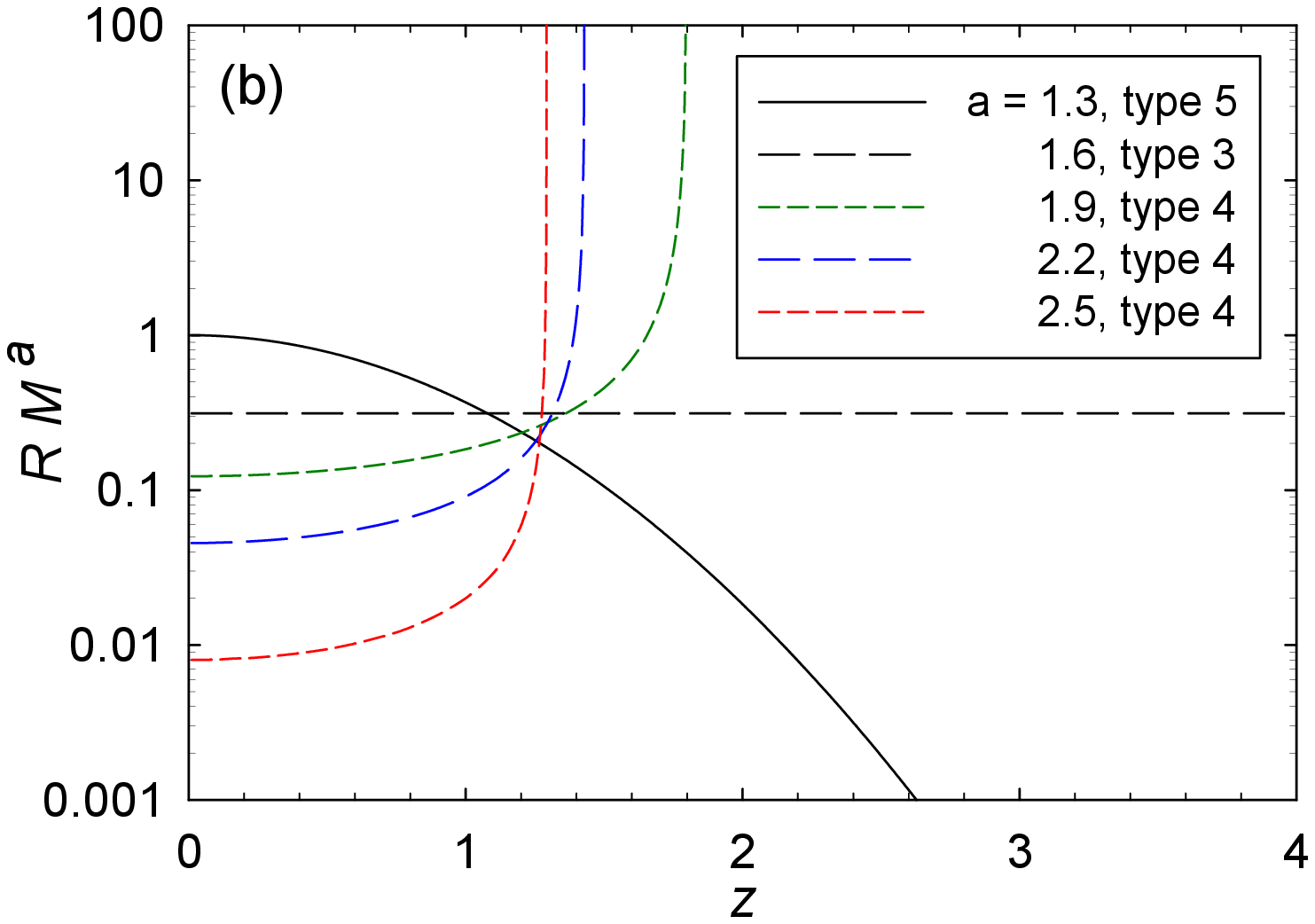}
\end{minipage}
\caption{Graphs for $T$ and $R$ with $b=1.3$ and increasing values of $a$. We see instances of types 3, 4, 5, and 6. Solutions for $Y(z)$ cease to be analytic when $R$ diverges to either $+\infty$ or $-\infty$ (the type 6 divergence of $R$ is not shown on the log scale since $R$ is negative). These infinities in $|R|$ mark points of termination, indicated on the $T$ graphs by bold dots. Type $3$ and $5$ curves do not terminate.}
\label{fig:Figure6}
\end{figure}

Figure \ref{fig:Figure6} plots $T$ and $R$ for $b=1.3$ and several values of $a$. Solutions for $Y(z)$ become nonanalytic where $|R|$ diverges. I did not extend my solutions beyond these points of non-analyticity, since there is too little guidance from the CCM for doing so. Types 3 and 5 have no diverging $R$ values, and value of $T$ and $R$ rapidly approach zero as $z\to\infty$.

\subsection{Final remarks about the GE solutions}

The GE solutions have produced candidate BHT equations of state in qualitative agreement with a number of features of the CCM. All of the GE solutions in my search grid have the entropy $S$ decreasing monotonically from maxima at $z=0$. With appropriate values of $(a,b)$, there are a number of GE solutions of types $4-6$ with $C_X$ starting negative, diverging to negative infinity, and then returning from positive infinity as $z$ increases. GE solutions of all types except type $1$ have metric determinants $p_2$ uniformly negative. GE solutions of types $4$ and $5$ have $T$ decreasing monotonically from maxima at $z=0$. Type $4$ has positive values of $R$ diverging to infinity, matching qualitatively a basic feature of Kerr in the CCM. These are all encouraging results, particularly considering that just the two exponents $(a,b)$ were varied.

\par
Areas of difference concern the character of the BHT extremal point. In the CCM, general relativity has $T\to 0$ correspond to zero surface gravity, and, at least for Kerr, $R\to\infty$. As was pointed out by Ruppeiner \cite{Ruppeiner2008}, this extremal behavior for Kerr has features in common with the ideal Fermi gas. But the CCM entropies do not go to zero in the extremal limit, inconsistent with usual statements \cite{Pathria2011} of the third law of thermodynamics that has $S\to 0$ as $T\to 0$. This discrepancy results from the event horizon area (proportional to $S$) not going to zero in this limit.

\par
The GE solutions show a somewhat different extremal limit. $S$ decreases monotonically to zero, but the zero coincides neither with any zero in $T$ nor to a divergence in $R$. Although the position of this entropy zero might be adjustable, the CCM gives no particular insight about how this should be done. Therefore, I will not pursue this further here. Generally, types $4$ and $5$ seem physically the most interesting.

\section{Model comparisons}

The main agenda of this paper is to get the BHT equations of state by thermodynamic means. But there is another method of proceeding: BHT solutions derived from known black hole models can be tested for agreement with the GE. A detailed exploration of this theme is beyond the scope of this manuscript, but this section presents a brief review of the large recent literature of BHT papers computing the thermodynamic curvature $R$ from various classical and quantum models. An example of comparing model solutions with the GE was by Sahay and Jha \cite{Sahay2017b} who demonstrated that near the phase transition in the 5d Gauss-Bonnet-AdS black hole, the free energy expression satisfies the GE.

\par
In the general relativistic scenario, most of the recent models have added AdS backgrounds. AdS backgrounds bring thermodynamic stability, Hawking-Page phase transitions, and connections to AdS/CFT correspondence. These classical scenarios have seen extensions of spatial dimensionality via the Myers-Perry black holes \cite{Aman2006a, Aman2013}, and via exotic local topologies \cite{Zhang2015, Arcioni2005}. The issue of the appropriate number of independent variables to use has been raised \cite{ Mirza2007}. Classical RN-AdS black holes have van der Waals phase transitions \cite{Chamblin1999,Shen2007, Sahay2010b, Niu2012, Sahay2010a}. Attention has been paid to adding the pressure as a thermodynamic variable, with the conjugate variable being the volume enclosed by the black hole \cite{Mann2012, Dolan2015}.

\par
There are a number of string theory approaches starting from microscopics, and employing state-counting schemes to directly compute the entropy \cite{Strominger1996}. Such attempts using the geometry of thermodynamic include \cite{Sarkar2008, Bellucci2010, Bellucci2012, Wei2015, Sengupta2017a}. There are also string theory models based on novel particles, such as R-charges \cite{Sahay2010c}, dyons \cite{Sengupta2017b}, and tidal charges \cite{Pidokrajt2011a}.

\par
Also on the scene is a broad class of models appending terms quadratic and higher-order in the space-time Ricci curvature scalar to the Einstein-Hilbert Lagrangian. These approaches usually also employ Maxwell terms, and contain free parameters calculable by string theory. Such models include Gauss-Bonnet AdS gravity \cite{Sahay2017b,Wei2013}, Ho\v{r}ava-Lifshitz gravity \cite{Wei2012, Bagher2017}, $f(R)$ gravity \cite{Deng2017, Li2016}, and Einstein-dilaton-Lifshitz gravity \cite{Zangeneh2017}. Another class of black holes to which geometry of thermodynamics has been applied employs nonlinear electrodynamics (NED) \cite{Wei2018}. There have also been efforts to model BHT phase transitions with Ehrenfest's theory of phase transitions \cite{Banerjee2012, Lala2012}. In addition, some authors have used different thermodynamic metrics in the black hole regime \cite{Quevedo2008, Hendi2017}.

\section{Conclusion}

In this paper, I have argued that fundamental properties about black hole thermodynamics might be obtained by thermodynamic means. General assumptions about the number of conserved thermodynamic variables, conditions of analyticity, and limiting cases, yield a scaled thermodynamics as a solution to an ordinary differential equation parameterized by two exponents $(a,b)$. The basic agenda proposed here is that information about black holes from either models or from astrophysical observations could serve to constrain $(a,b)$ and correspondingly improve our knowledge of black hole thermodynamics in real physical scenarios.
 
\par
Because the true black hole thermodynamic equations of state are so hard to get, in the absence of a consensus microscopic model, the method here is hard to test directly. However, the same method may be employed in ordinary thermodynamics, where there is a good grounding on statistical mechanics. Several successful tests of my method have been made in this conventional scenario. If we believe in the unity of thermodynamics, there may be reason to think that the method proposed here will be effective in a broader scenario, including black holes.

\section{Acknowledgements}

I thank Anurag Sahay for useful conversations at the SigmaPhi 2017 conference, and Alex Roman for locating several references.

\section{Appendix: general formulas}

In this section, I list formulas and series for several quantities. The series are generated starting with

\begin{equation} Y(z)=y_0+y_2 z^2 + y_4 z^4 + O(z^6),\label{400}\end{equation}

\noindent where

\begin{equation} z=\frac{X}{M^b}.\label{410}\end{equation}

\noindent In this paper, I consider without loss of generality only $y_0=1$ and $y_2=-1$. The entropy is

\begin{equation} S M^{-a}= Y(z), \label{420}\end{equation}

\noindent with 

\begin{equation} S M^{-a}=y_0+y_2 z^2 +O(z^4). \label{430}\end{equation}

\noindent The temperature is

\begin{equation} T M^{a-1}=\frac{1}{a Y(z)-b z Y'(z)}, \label{440} \end{equation}

\noindent with

\begin{equation} T M^{a-1}=\frac{1}{a y_0}-\frac{y_2 (a-2 b)}{a^2 y_0^2}z^2 + O\left(z^4\right).\label{450}\end{equation}

\noindent The free energy is

\begin{equation} \phi M^{-a}= \left[(b-1) z Y'(z)-(a-1) Y(z)\right], \label{460}\end{equation}

\noindent with

\begin{equation} \phi M^{-a}=-(a-1) y_0 -(1 + a - 2 b) y_2 z^2 +O\left(z^4\right). \label{470}\end{equation}

\noindent The heat capacity at constant $X$ is

\begin{equation}C_X M^{-a}=-\frac{\left[a Y(z)-b z Y'(z)\right]^2}{b z \left[(-2 a+b+1) Y'(z)+b z Y''(z)\right]+(a-1) a Y(z)},\label{480}\end{equation}

\noindent with

\begin{equation}C_X M^{-a}= -\frac{a y_0}{a-1}-\frac{y_2 \left[a^2-a+2 (1-2 b) b\right]}{(a-1)^2}z^2+O\left(z^4\right). \label{490}\end{equation}

\noindent The first determinant coefficient is

\begin{equation} p_1 M^{2-a}=a(1-a) Y(z)-b z \left[(1-2 a+b) Y'(z)+b z Y''(z)\right], \label{500}\end{equation}

\noindent with

\begin{equation} p_1 M^{2-a}=-a(a-1) y_0+y_2\left(-a^2+4 a b+a-4 b^2 - 2b\right)z^2 + O\left(z^4\right). \label{510}\end{equation}

\noindent The second determinant coefficient is

\begin{equation} p_2 M^{2 - 2 a + 2 b}=-(a-b)^2 Y'(z)^2+a(a-1) Y(z) Y''(z)-b(b-1) z Y'(z) Y''(z), \label{520}\end{equation}

\noindent with

\begin{multline} p_2 M^{2 - 2 a + 2 b}=2a (a-1) y_0 y_2-\\\,\,\,\,\,\,\,\,\,2 \left(a^2 y_2^2-4 a b y_2^2+a y_2^2+4 b^2 y_2^2 -2b y_2^2 + 6 a y_0 y_4 - 6 a^2 y_0 y_4\right)z^2+O\left(z^4\right). \label{530}\end{multline}

\noindent The thermodynamic curvature is

\begin{equation} R M^{a}=\frac{\mathcal N_1}{\mathcal D_1}, \label{540}\end{equation}

\noindent where

\begin{multline}
\mathcal N_1=-(b-1)\times
 \\
 \!\!\!\!\!\!\!\!\!\!\!\! \Big\{-Y'(z) [b z (a^2-3 a b+2 b) Y''(z)^2+(a-1) a (a-b) Y(z) Y^{(3)}(z)-(b-1) b^2 z^2 Y^{(3)}(z) Y''(z)] +\\
\!\!\!\! (b-a) Y'(z)^2 [(a^2-2 a b-a+b^2+b) Y''(z)+b z (-2 a+b+1) Y^{(3)}(z)]+\\
(a-1) a Y(z) Y''(z)[2 (a-2 b) Y''(z)-b z Y^{(3)}(z)]-(b-1) b^2 z^2 Y''(z)^3\Big\},
\label{550}
\end{multline}

\noindent and

\begin{equation} \mathcal D_1 = 2 \left[(a-b)^2 Y'(z)^2-(a-1) a Y(z) Y''(z)+(b-1) b z Y'(z) Y''(z)\right]^2. \label{560}\end{equation}

\noindent The series for $R$ is given in Eq. (\ref{260}):

\begin{equation} \displaystyle R M^a=-\frac{(b-1) (a-2 b)}{(a-1) a y_0}\,+\frac{\mathcal N_2}{(a-1)^2 a^2 y_0^2 y_2} z^2+O\left(z^4\right),\label{570}\end{equation}

\noindent with

\begin{multline}
\mathcal N_2=2 (b-1)\times
 \\
\!\!\!\!\!\!\!\!\!\!\!\! \Big\{-a^3 \left(y_2^2-3 y_0 y_4\right)+a^2 \left[(6 b-1) y_2^2-3 y_0 y_4\right]+4 a (1-3 b) b y_2^2+4 b^2 (2 b-1) y_2^2\Big\}.
\end{multline}

 \end{document}